\documentclass[12pt,letterpaper]{article}

\usepackage{amsmath,amssymb,amsthm,bm}
\newtheorem{lemma}{Lemma}
\newtheorem{proposition}{Proposition}
\newtheorem{corollary}{Corollary}
\newtheorem{theorem}{Theorem}
\usepackage{graphicx}
\usepackage{booktabs}
\usepackage{comment}
\usepackage{url}
\usepackage{placeins}
\usepackage{tikz}
\usetikzlibrary{calc,arrows.meta}
\usepackage[ruled,vlined]{algorithm2e}

\usepackage[authoryear,round]{natbib}
\usepackage[
    colorlinks=true,
    linkcolor=black,
    citecolor=blue,
    urlcolor=blue
]{hyperref}

\newcommand{\blind}{0}

\newcommand{\papertitle}{%
Filtration-Based Orthogonalized Moment Aberration for General Treatment Designs with Structured Experimental Units}

\begin{document}

\def\spacingset#1{\renewcommand{\baselinestretch}{#1}\small\normalsize}
\spacingset{1}

\ifnum\blind=0
  \title{\bfseries\papertitle}
\author{%
Ming-Chung Chang$^{1}$\thanks{Corresponding author.
Email: \texttt{mcchang0131@as.edu.tw}.},
\quad Yun-Hsuan Lien$^{2}$,
\quad and Che-Shi Lee$^{2}$\\[3pt]
{\small $^{1}$Institute of Statistical Science, Academia Sinica}\\
{\small $^{2}$Research Center for Information Technology Innovation,
Academia Sinica}\\
{\small Taipei, Taiwan}
}
\date{}
\date{}
  \date{}
  \maketitle
\else
  \begin{center}
    {\LARGE\bfseries\papertitle\par}
  \end{center}
\fi

\thispagestyle{empty} 


\begingroup
\renewcommand{\baselinestretch}{1}
\normalsize
\maketitle
\thispagestyle{empty}
\vspace{-2em}

\begin{abstract}
Modern experiments often involve complex treatment spaces and heterogeneous
experimental units. We propose filtration-based orthogonalized moment aberration
(F-OMA), a model-free framework for constructing general treatment designs.
First, F-OMA accommodates diverse treatment filtrations, including
effect-hierarchy, geometric and partition-based, and spectral and multiresolution
filtrations, recovering classical aberration and certain space-filling criteria
as special cases. Second, independently specified positive semidefinite unit
kernels accommodate nested and crossed block structures, partially-relaxed
orthogonal block structures (PR-OBS), and more general experimental-unit
relationships. Third, we characterize simultaneous optimality under unknown
PR-OBS stratum weights and establish statistical interpretations through treatment--unit
interaction variance, kernel discrepancies, and a layer-weighted prediction-error
bound under additional assumptions. We also develop a constrained multistart local-search algorithm,
implemented in the \texttt{FOMA} R package,
for practical design construction. Numerical studies compare F-OMA
with alternative initializers for Bayesian optimization under nested and crossed
unit structures, using unit-aware and unit-blind Gaussian process surrogates.
Together, these developments extend aberration-based experimental design to
more general treatment and unit structures.

\end{abstract}
\endgroup
\clearpage

\spacingset{2} 

\section{Introduction}
\label{sec:introduction}

Experimental design has long played an important
role in agriculture and industry \citep{fisher1926,wuhamada2021}.
Representative approaches include factorial
designs \citep{cheng2014,chengtang2025},
response surface designs \citep{boxwilson1951},
and space-filling designs
\citep{mckay1979,joseph2015}.
Experimental design is also important in artificial
intelligence (AI), where the hyperparameters of a training algorithm can be viewed as the treatment factors in an experiment,
and validation loss or accuracy as the observed response
\citep{snoek2012}.
Moreover, training performance can vary
across data splits and random initializations
\citep{bouthillier2021}.
When these conditions form structured experimental units,
design quality depends not only on which configurations
are selected but also on how they are assigned to these
units.
This is exactly a central consideration in multi-stratum designs,
where experimental units have certain \emph{block structures} \citep{nelder1965,bailey2008,cheng2014}.

Existing multi-stratum designs have
primarily focused on factorial treatments
\citep{chengtsai2011,changcheng2018,chang2025}
and polynomial response surfaces
\citep{trincagilmour2001,trincagilmour2026},
with limited consideration of more general
treatment-function
spaces.
Separately, space-filling designs emphasize
geometric coverage of quantitative treatment
regions, but conventional space-filling
criteria typically do not explicitly account
for structured experimental units
\citep{santner2018}.
These complementary perspectives motivate
a unified framework that independently
specifies treatment-function spaces and
experimental-unit structures.
\citet{chang2025} introduced orthogonalized moment
aberration (OMA) for multi-stratum factorial designs,
but its treatment-function
spaces rely on orthogonalized
pointwise powers of a treatment kernel, while its
unit kernels are derived from block structures.
In this paper, we propose
\emph{filtration-based orthogonalized moment aberration}
(\emph{F-OMA}), a model-free design framework
that specifies treatment priorities through
nested treatment-function spaces and accommodates
arbitrary positive semidefinite unit kernels,
extending OMA beyond factorial treatments
and block structures.

Our main methodological contributions are fourfold.
First, F-OMA provides a model-free design framework
accommodating diverse treatment filtrations,
including effect-hierarchy, geometric and
partition-based, and spectral and multiresolution
filtrations.
With suitable choices, F-OMA recovers classical
aberration and certain space-filling criteria
while allowing more general treatment-function
hierarchies.
Second, independently specified positive
semidefinite unit kernels accommodate nested
and crossed block structures, partially-relaxed
orthogonal block structures (PR-OBS)
\citep{chang2025}, and more general unit
relationships.
Using the established PR-OBS stratum
decomposition, we characterize simultaneous
optimality under unknown stratum weights.
Third, we establish statistical interpretations
of F-OMA scores through treatment--unit
interaction variance and maximum mean
discrepancy (MMD).
Under additional modeling assumptions,
we also derive a layer-weighted generalized
least squares (GLS) prediction-error bound
involving F-OMA scores and sampled-feature
coverage.
Fourth, we develop a constrained multistart
local search using margin-preserving configuration
exchanges and unit-assignment swaps.
The accompanying \texttt{FOMA} R package supports
the construction and evaluation of designs under
diverse treatment filtrations and unit kernels.
We illustrate the applicability of F-OMA through
Bayesian optimization simulations under nested and
crossed unit structures, comparing different initial
designs using unit-aware and unit-blind Gaussian
process surrogates.

The remainder of this paper is organized
as follows.
Section~\ref{sec:framework} introduces the
F-OMA framework.
Section~\ref{sec:treatment-filtrations}
presents representative treatment filtrations
and their connections to existing criteria.
Section~\ref{sec:theory} establishes theoretical
properties and statistical interpretations.
Section~\ref{sec:numerical-studies} presents numerical
studies under nested and crossed unit structures.
Section~\ref{sec:discussion} summarizes the contributions,
findings, and scope of applicability.
The appendices contain proofs, additional theoretical results,
a constrained design construction algorithm, simulation details,
and supplementary numerical results.

\section{A General Framework for F-OMA}
\label{sec:framework}

We first introduce the design setting and then
define the F-OMA criterion by combining two
independently specified components:
a \emph{treatment filtration} that prioritizes
treatment features and a \emph{unit kernel}
that represents relationships among
experimental units.

\subsection{Treatment Assignments and Filtrations}
\label{subsec:design-setup}

Let
$\mathcal{X}=\{\mathbf{x}^{(1)},\ldots,\mathbf{x}^{(n)}\}$
be a finite set of candidate treatments and
$\Omega=\{u_1,\ldots,u_N\}$ the set of experimental units.
An $N$-run design is a map $d:\Omega\to\mathcal{X}$,
where $d(u)=\mathbf{x}$ assigns treatment
$\mathbf{x}\in\mathcal{X}$ to unit $u\in\Omega$.
With the unit order fixed, we write
$d=(\mathbf{x}_1,\ldots,\mathbf{x}_N)
\in\mathcal{D}\subseteq\mathcal{X}^N$,
where $\mathcal{D}$ denotes the set of feasible assignments.
The set of selected treatments is
$D(d)=d(\Omega)=\{\mathbf{x}_1,\ldots,\mathbf{x}_N\}
\subseteq\mathcal{X}$.
Define the assignment matrix
$\mathbf{X}_d\in\{0,1\}^{N\times n}$ by
$(\mathbf{X}_d)_{ij}
=\mathbb{I}\{d(u_i)=\mathbf{x}^{(j)}\}$,
where $\mathbb{I}\{\cdot\}$ is the indicator function.
Each row sums to one, and
$\mathbf{r}_d=\mathbf{X}_d^\top\mathbf{1}_N$
gives the treatment replication counts,
where $\mathbf{1}_N$ is the $N$-vector of ones.

Scientific investigations often begin by
identifying broad patterns before examining
finer details.
With limited experimental resources,
this motivates prioritizing simpler or
coarser treatment features before more
complex or finer ones.
We formalize these priorities through
nested treatment-function spaces.
Let $\mathcal H=\mathbb R^n$ be the space
of real-valued functions on $\mathcal X$,
equipped with the standard Euclidean inner
product unless otherwise specified.
Section~\ref{subsec:geometric-filtrations}
extends the partition-based construction
to general reference probability measures.
A treatment filtration is a sequence of nested
subspaces
\[
\mathcal{H}_0
\subset \mathcal{H}_1
\subset \cdots
\subset \mathcal{H}_T
\subseteq \mathcal{H},
\]
where
$\mathcal{H}_0=\operatorname{span}\{\mathbf{1}_n\}$.
The filtration is specified before experimentation
and determines the priority order of treatment features.
Let $\mathbf{P}_t$ denote the orthogonal projector
onto $\mathcal{H}_t$.
For $t=1,\ldots,T$, define the orthogonal
increment space
\(
\mathcal{S}_t
=
\mathcal{H}_t\ominus\mathcal{H}_{t-1},
\)
with the corresponding projector
\(
\mathbf{B}_t
=
\mathbf{P}_t-\mathbf{P}_{t-1}.
\)
The increment spaces are mutually orthogonal,
and any $\mathbf{f}\in\mathcal{H}$ admits the
decomposition
\(
\mathbf{f}
=
\mathbf{P}_0\mathbf{f}
+
\sum_{t=1}^{T}\mathbf{B}_t\mathbf{f}
+
(\mathbf{I}_n-\mathbf{P}_T)\mathbf{f}.
\)
Thus, each increment contains features not
included in earlier layers.
Let $\mathbf{Q}_t\in\mathbb{R}^{n\times r_t}$
have orthonormal columns spanning
$\mathcal{S}_t$, where
$r_t=\dim(\mathcal{S}_t)$.
Then
\(
\mathbf{B}_t=\mathbf{Q}_t\mathbf{Q}_t^\top.
\)
For a design $d$, define the treatment feature
matrix at layer $t$ as
$\mathbf{Z}_t(d)=\mathbf{X}_d\mathbf{Q}_t
\in\mathbb{R}^{N\times r_t}$.
Its $N$ rows represent the treatment features
assigned to the $N$ experimental units at that layer.
These matrices are used to define the F-OMA criterion.

We note that OMA \citep{chang2025} is recovered by choosing
$\mathcal H_t=\operatorname{span}\{\mathbf 1_n\}
+\sum_{k=1}^{t}\operatorname{col}(\mathbf G^{\circ k})$
for a treatment-kernel matrix $\mathbf G$, where
$\mathbf G^{\circ k}$ denotes its $k$th entrywise power
and $\operatorname{col}$ denotes column space.
The extension is strict. For example, on
$\mathcal{X}=\{1,2,3,4\}$, consider
\[
\mathcal{H}_0=\operatorname{span}\{\mathbf{1}_4\},
\qquad
\mathcal{H}_1=\{(a,a,b,b)^\top:a,b\in\mathbb{R}\},
\qquad
\mathcal{H}_2=\mathbb{R}^4.
\]
The first increment compares the groups $\{1,2\}$ and
$\{3,4\}$, while the second distinguishes treatments
within each group.
Any kernel matrix $\mathbf{G}$ satisfying
$\operatorname{span}\{\mathbf{1}_4\}
+\operatorname{col}(\mathbf{G})=\mathcal{H}_1$
has identical rows within each group.
Entrywise powers preserve these equalities, so the
corresponding OMA hierarchy remains at $\mathcal{H}_1$.
F-OMA accommodates the second increment by specifying
$\mathcal{H}_2$ directly.

\subsection{Unit Kernels and Structured Experimental Units}
\label{subsec:unit-kernels}

We refer to the relationships among experimental
units as the unit structure and represent them
by a symmetric positive semidefinite matrix
$\mathbf V\in\mathbb R^{N\times N}$,
called the \emph{unit kernel}.
Positive semidefiniteness ensures that
$(\mathbf V)_{ij}=\boldsymbol\phi(u_i)^\top
\boldsymbol\phi(u_j)$ for some feature map
$\boldsymbol\phi$, allowing the kernel to measure
similarity between experimental units.
It may represent block structures or more
general unit relationships.
Let the spectral decomposition be
$\mathbf V=\sum_{r=1}^N\eta_r
\mathbf u_r\mathbf u_r^\top$,
where $\eta_r\ge0$ and
$\mathbf u_1,\ldots,\mathbf u_N$
are orthonormal eigenvectors.
Each eigenvector represents a unit-kernel
eigenmode, weighted by its eigenvalue.

In many applications, the unit structure
can be described by a collection of \emph{unit factors},
each partitioning the units into blocks.
Such a unit structure is referred to as a
block structure.
Let $\mathcal F$ denote the collection of unit factors,
and let $\mathbf P_F$ be the orthogonal projector
onto the space of vectors that are constant within
each block of factor $F\in\mathcal F$.
Given a block structure, a unit kernel can then be constructed as
$\mathbf V=\sum_{F\in\mathcal F}w_F\mathbf P_F$,
where $w_F\geq 0$, accommodating orthogonal
block structures
\citep{bailey1996}
and PR-OBS \citep{chang2025}.
More generally, F-OMA allows any positive
semidefinite unit kernel, including those
reflecting unit dependence from chains of
randomizations \citep{baileybrien2016}
or association schemes \citep{bailey2004},
without requiring a block structure.

\subsection{The F-OMA Criterion and Basic Properties}
\label{subsec:foma-criterion}

With limited experimental resources,
treatments with similar features may provide
redundant information, particularly when
assigned to structurally related units.
To account for both treatment selection and
unit assignment, we define a \emph{treatment kernel}
for each filtration layer and combine it with
the unit kernel to construct the F-OMA criterion.

For each filtration layer $t$, define
the treatment kernel, also called the
\emph{layer kernel}, by
\begin{equation*}
p_t(\mathbf{x}^{(a)},\mathbf{x}^{(b)})
=
(\mathbf{B}_t)_{ab}
=
\sum_{\ell=1}^{r_t}
(\mathbf{Q}_t)_{a\ell}(\mathbf{Q}_t)_{b\ell},
\end{equation*}
where $\mathbf{B}_t$ and $\mathbf{Q}_t$ are defined
in Section~\ref{subsec:design-setup}.
The kernel value $p_t(\mathbf{x}^{(a)},\mathbf{x}^{(b)})$
measures the similarity between treatments
$\mathbf{x}^{(a)}$ and $\mathbf{x}^{(b)}$ in the
$t$th filtration layer.
Given a design $d$ and a unit kernel
$\mathbf{V}=(V_{ij})\succeq 0$,
define the F-OMA score at layer $t$ by
\begin{equation}
\label{eq:foma-layer}
K_t(d;\mathbf{V})
=
\frac{1}{N^2}
\sum_{i=1}^{N}\sum_{j=1}^{N}
p_t(\mathbf{x}_i,\mathbf{x}_j)V_{ij},
\qquad t=1,\ldots,T.
\end{equation}
Thus, F-OMA measures pairwise treatment-feature
similarities, weighted by the unit kernel.
When the unit kernel assigns larger positive weights
to more homogeneous unit pairs, greater similarity
in their treatment features at layer $t$
tends to increase $K_t(d;\mathbf{V})$.

Let
\(
\mathbf{H}_t(d)
=
\bigl(p_t(\mathbf{x}_i,\mathbf{x}_j)\bigr)_{i,j=1}^{N}.
\)
By the definitions in Section~\ref{subsec:design-setup},
\(
\mathbf{H}_t(d)
=
\mathbf{X}_d\mathbf{B}_t\mathbf{X}_d^\top
=
\mathbf{Z}_t(d)\mathbf{Z}_t(d)^\top.
\)
Therefore, the F-OMA score has the equivalent
trace representation
\begin{equation}
\label{eq:foma-projector}
K_t(d;\mathbf V)
=\frac{1}{N^2}
\operatorname{tr}\{\mathbf V\mathbf H_t(d)\}
=\frac{1}{N^2}
\operatorname{tr}\{
\mathbf Z_t(d)^\top
\mathbf V\mathbf Z_t(d)\}.
\end{equation}
This representation shows that the F-OMA score
depends only on the treatment increment space
and the unit kernel, not on the choice of
orthonormal basis.
Since $\mathbf V\succeq0$, its spectral
decomposition gives
$K_t(d;\mathbf V)
=N^{-2}\sum_{r=1}^N\eta_r
\|\mathbf Z_t(d)^\top\mathbf u_r\|^2\ge0$,
where $\eta_r$ and $\mathbf u_r$ are the
eigenvalues and orthonormal eigenvectors of
$\mathbf V$.
Thus, $K_t(d;\mathbf V)$ measures the alignment
between treatment features and unit-kernel
eigenmodes.
When $\mathbf V=\mathbf 1_N\mathbf 1_N^\top$,
we have
$K_t(d;\mathbf V)
=N^{-2}\|\mathbf Q_t^\top\mathbf r_d\|^2$,
which depends only on treatment replication
counts and recovers classical minimum
aberration criteria under suitable filtrations
(Section~\ref{sec:treatment-filtrations}).

The orthogonal increments of the treatment
filtration separate new treatment features
from those represented at earlier layers.
To prioritize these features while accounting
for experimental-unit relationships,
we define the F-OMA criterion as
$\mathbf K(d;\mathbf V)
=(K_1(d;\mathbf V),\ldots,K_T(d;\mathbf V))$.
A design $d^\star\in\mathcal D$ is
F-OMA-optimal if
$\mathbf K(d^\star;\mathbf V)
\preceq_{\mathrm{lex}}\mathbf K(d;\mathbf V)$
for all $d\in\mathcal D$,
where $\preceq_{\mathrm{lex}}$ denotes
lexicographic ordering, giving priority
to earlier filtration layers.
For scalar optimization, consider
$A_{\mathbf a}(d;\mathbf V)
=\sum_{t=1}^T a_tK_t(d;\mathbf V)$,
where $a_t>0$.
The following result shows that suitable weights
preserve the lexicographic ordering, allowing standard scalar optimization methods
to be used while preserving the priority of
earlier filtration layers.
The proof is given in Appendix \ref{app:proof-foma-scalarization}.
\begin{proposition}
\label{prop:foma-scalarization}
There exists $\epsilon_0>0$ such that,
for every $0<\epsilon<\epsilon_0$,
the scalar criterion $A_{\mathbf{a}}(d;\mathbf{V})$
with $a_t=\epsilon^{t-1}$ gives exactly
the same ordering of designs as
the lexicographic F-OMA criterion.
\end{proposition}

For practical construction,
Appendix~\ref{app:construction} presents
a multistart local search
(Algorithm~\ref{alg:foma-exchange})
for F-OMA designs with distinct treatments
and exact one-factor marginal balance.
It uses margin-preserving factor-level
exchanges and unit-assignment swaps
with exact score updates.
The \texttt{FOMA} R package implements this algorithm
and provides tools for specifying treatment filtrations
and unit kernels, constructing designs, and evaluating
F-OMA scores.

\section{Treatment Filtrations and Connections to Existing Design Criteria}
\label{sec:treatment-filtrations}

F-OMA allows treatment filtrations to
be specified directly according to design objectives.
We consider three classes of treatment
filtrations: \emph{effect-hierarchy},
\emph{geometric and partition-based}, and
\emph{spectral and multiresolution}.
Suitable choices recover established
design criteria, while each filtration
can be combined with any positive
semidefinite unit kernel through
\eqref{eq:foma-layer}.
Throughout this paper, $|\cdot|$ denotes
cardinality for sets, absolute value or
modulus for scalars, and total degree for
multi-indices, depending on the argument.

\subsection{Effect-Hierarchy Filtrations}
\label{subsec:effect-hierarchy}

Effect-hierarchy filtrations prioritize treatment
features by interaction order or polynomial degree.
These hierarchies are commonly used in factorial
and response-surface designs
\citep{chang2022,chengye2004}.

\paragraph*{Factorial-order filtration.}

Consider $m$ treatment factors, where
$\mathcal X_j$ denotes the set of $s_j$ levels
of the $j$th factor.
The full factorial candidate space is
$\mathcal X=\mathcal X_1\times\cdots\times\mathcal X_m$,
with $n=\prod_{j=1}^m s_j$ treatment combinations.
Let $\mathcal C_j=\{\mathbf v\in\mathbb R^{s_j}:
\mathbf v^\top\mathbf 1_{s_j}=0\}$
denote the contrast space of the
$j$th treatment factor.
For $A\subseteq\{1,\ldots,m\}$, define
$\mathcal{W}_A=\bigotimes_{j=1}^m\mathcal{E}_{j,A}$,
where $\mathcal{E}_{j,A}=\mathcal{C}_j$ if $j\in A$
and $\mathcal{E}_{j,A}
=\operatorname{span}\{\mathbf{1}_{s_j}\}$ otherwise.
The factorial-order filtration is
$\mathcal{H}_t^{\mathrm{fac}}
=\bigoplus_{|A|\leq t}\mathcal{W}_A$,
with orthogonal increments
$\mathcal{S}_t^{\mathrm{fac}}
=\bigoplus_{|A|=t}\mathcal{W}_A$,
for $t=1,\ldots,m$.
Here $\bigotimes$ denotes the tensor product of vector spaces,
and $\bigoplus$ denotes an orthogonal direct sum.
Thus, the filtration gives priority to main effects,
followed by two-factor interactions and higher-order
interactions, as in classical minimum aberration
\citep{fries1980,xuwu2001}.

\paragraph*{Polynomial-degree filtration.}

For quantitative factors, polynomial degree provides
another treatment hierarchy.
Let $\rho_{j,0},\ldots,\rho_{j,s_j-1}$
be orthonormal polynomial functions on
$\mathcal X_j$ under the discrete inner product
$\langle f,g\rangle_j
=\sum_{x\in\mathcal X_j}f(x)g(x)$,
with $\rho_{j,0}(x)=s_j^{-1/2}$
for all $x\in\mathcal X_j$.
For a multi-index
$\boldsymbol{\alpha}=(\alpha_1,\ldots,\alpha_m)$,
define $\rho_{\boldsymbol{\alpha}}(\mathbf{x})
=\prod_{j=1}^m\rho_{j,\alpha_j}(x_j)$
and $|\boldsymbol{\alpha}|=\sum_{j=1}^m\alpha_j$.
The polynomial-degree filtration is
$\mathcal{H}_t^{\mathrm{poly}}
=\operatorname{span}\{\rho_{\boldsymbol{\alpha}}:
|\boldsymbol{\alpha}|\leq t\}$,
with orthogonal increments
$\mathcal{S}_t^{\mathrm{poly}}
=\operatorname{span}\{\rho_{\boldsymbol{\alpha}}:
|\boldsymbol{\alpha}|=t\}$,
for $t=1,\ldots,\sum_{j=1}^m(s_j-1)$.
This filtration prioritizes lower-degree
polynomial effects, as in the
$\beta$-aberration criterion of
\citet{chengye2004}.

The following result connects these filtrations
to established aberration criteria.
The proof is given in
Appendix~\ref{app:classical-aberration}.
\begin{proposition}
\label{prop:classical-aberration}

Let $K_t^{\mathrm{fac}}(d;\mathbf V)$ and
$K_t^{\mathrm{poly}}(d;\mathbf V)$ denote
the F-OMA layer scores corresponding to the
factorial-order and polynomial-degree filtrations,
respectively.
For $\mathbf{V}=\mathbf J_N=\mathbf1_N\mathbf1_N^\top$,
we have
$K_t^{\mathrm{fac}}(d;\mathbf J_N)=A_t(d)/n$,
$t=1,\ldots,m$, where $A_t(d)$ is the
generalized wordlength component of
\citet{xuwu2001}.
Similarly,
$K_t^{\mathrm{poly}}(d;\mathbf J_N)=\beta_t(d)/n$,
$t=1,\ldots,\sum_{j=1}^m(s_j-1)$,
where $\beta_t(d)$ is the $\beta$-aberration
component of \citet{chengye2004}.
\end{proposition}

Mixed qualitative and quantitative factors
can be accommodated using contrast spaces
and orthogonal polynomial contrasts, respectively.
Their tensor products define treatment effects
ordered by interaction order and polynomial
degree.

\subsection{Geometric and Partition-Based Filtrations}
\label{subsec:geometric-filtrations}

We consider two constructions: nested
partition filtrations and filtrations based
on rectangular stratification and coordinate
projections.
These constructions are related to generalized
Haar representations \citep{girardisweldens1997}
and stratified discrepancy criteria
\citep{tianxu2026}.
Further examples, including hexagonal,
Voronoi, and manifold-based filtrations,
are given in
Appendix~\ref{app:filtration-illustrations}.

\paragraph*{Nested partition filtrations.}

Let $\nu$ be a reference probability measure
on a finite or continuous treatment space
$\mathcal X$.
Consider finite nested measurable partitions
$\Pi_0,\ldots,\Pi_T$, where
$\Pi_0=\{\mathcal X\}$, each $\Pi_t$
refines $\Pi_{t-1}$, and $\nu(C)>0$
for every cell $C$.
Thus, $\sum_{C\in\Pi_t}\nu(C)=1$
at each level $t$.
Let $L^2(\nu)$ be the space of real-valued
square-integrable functions with inner product
$\langle f,g\rangle_\nu=\int fg\,d\nu$,
and let $\mathcal H_t$ be its subspace
of functions constant within each cell of
$\Pi_t$.
The orthogonal projector onto $\mathcal H_t$
has integral kernel
$P_t^\nu(\mathbf x,\mathbf x')
=\sum_{C\in\Pi_t}
\mathbb I\{\mathbf x\in C\}
\mathbb I\{\mathbf x'\in C\}/\nu(C)$.
The layer kernel
$p_t^\nu=P_t^\nu-P_{t-1}^\nu$,
$t=1,\ldots,T$, captures contrasts
introduced by each refinement.
For a design $d=(\mathbf x_1,\ldots,\mathbf x_N)$,
let $N_C(d)=\sum_{i=1}^N
\mathbb I\{\mathbf x_i\in C\}$
and define the cell-count discrepancy
\begin{equation}
\label{eq:partition-discrepancy}
\Delta_t(d;\nu)
=\sum_{C\in\Pi_t}
\frac{\{N_C(d)/N-\nu(C)\}^2}{\nu(C)},
\qquad \Delta_0(d;\nu)=0.
\end{equation}
Let $\mathbf e_C(d)
=(\mathbb I\{\mathbf x_i\in C\})_{i=1}^N$.
The following proposition, proved in
Appendix~\ref{app:proof-partition},
gives the corresponding F-OMA representations.

\begin{proposition}
\label{prop:partition-foma}
For any $\mathbf V\succeq0$, define
$K_t^\nu(d;\mathbf V)
=N^{-2}\sum_{i,j=1}^N
p_t^\nu(\mathbf x_i,\mathbf x_j)V_{ij}$.
For $\mathbf V=\mathbf J_N
=\mathbf1_N\mathbf1_N^\top$, we have
\begin{equation}
\label{eq:partition-foma-j}
K_t^\nu(d;\mathbf J_N)
=\Delta_t(d;\nu)-\Delta_{t-1}(d;\nu).
\end{equation}
More generally,
\begin{equation}
\label{eq:partition-general}
K_t^\nu(d;\mathbf V)
=\frac1{N^2}\left\{
\sum_{C\in\Pi_t}
\frac{\mathbf e_C(d)^\top
\mathbf V\mathbf e_C(d)}{\nu(C)}
-
\sum_{C\in\Pi_{t-1}}
\frac{\mathbf e_C(d)^\top
\mathbf V\mathbf e_C(d)}{\nu(C)}
\right\}.
\end{equation}
\end{proposition}

The above formulation uses the $L^2(\nu)$
inner product $\langle f,g\rangle_\nu=\int fg\,d\nu$.
For a uniform probability measure on a finite
candidate space, let $n=|\mathcal X|$ and
$n_C=|C|$, so that $\nu(C)=n_C/n$.
The vectors
$\mathbf q_C=n_C^{-1/2}
(\mathbb I\{\mathbf x^{(a)}\in C\})_{a=1}^n$,
$C\in\Pi_t$, form an orthonormal basis
of $\mathcal H_t$ under the Euclidean inner product.
Thus, the Euclidean orthogonal projector
$\mathbf P_t=\sum_{C\in\Pi_t}
\mathbf q_C\mathbf q_C^\top$ satisfies
$(\mathbf P_t)_{ab}
=n^{-1}P_t^\nu(\mathbf x^{(a)},\mathbf x^{(b)})$.
Consequently, both inner products lead to
the same F-OMA-optimal designs.

\paragraph*{Example 1: Nested triangular partitions.}

Consider $\mathcal X=[0,1]^2$ with
$\nu$ the normalized area measure, so that
$\nu(C)=\operatorname{Area}(C)$.
The square is first divided into two
equal-area triangles, each recursively
subdivided into four equal-area triangles
by connecting edge midpoints.
The resulting partitions
$\Pi_1,\Pi_2,\Pi_3$ contain 2, 8, and
32 cells, with increment dimensions
1, 6, and 24, respectively.

\begin{figure}[tbp]
\centering
\includegraphics[
width=.85\textwidth,
keepaspectratio
]{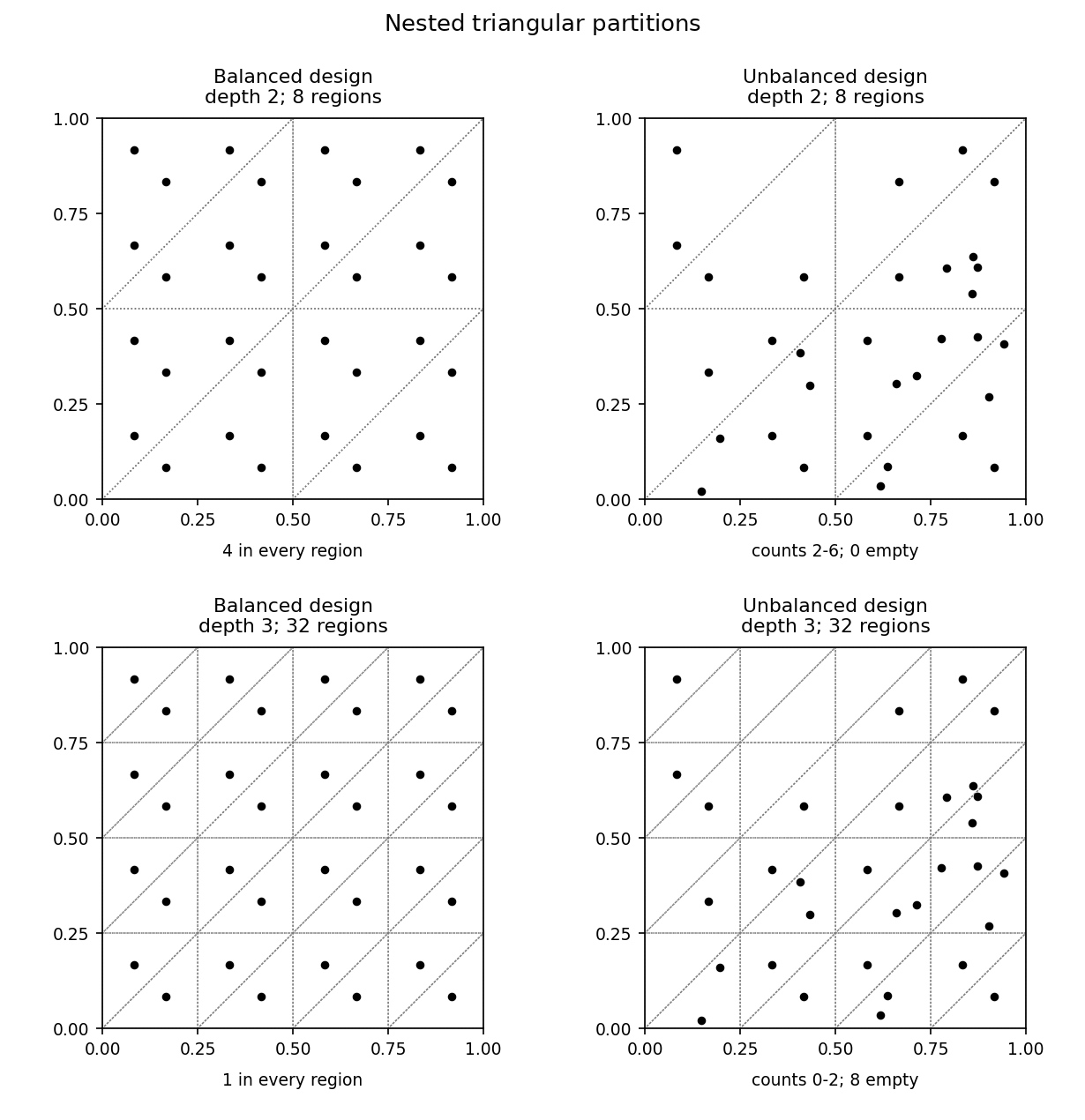}
\caption{Nested triangular partitions with
8 coarse cells and 32 refined cells,
comparing balanced and unbalanced
32-point designs.}
\label{fig:triangular}
\end{figure}

Figure~\ref{fig:triangular} compares two
32-point designs at depths two and three.
The balanced design has one point per
finest cell, giving zero discrepancy
at all depths.
For the unbalanced design,
$\Delta_1(d;\nu)=\Delta_2(d;\nu)=1/4$
and $\Delta_3(d;\nu)=1/2$.
By Proposition~\ref{prop:partition-foma},
their F-OMA score vectors under
$\mathbf V=\mathbf J_{32}$ are
$(0,0,0)$ and $(1/4,0,1/4)$, respectively.
The zero second-layer score illustrates
how orthogonal increments avoid counting
the same imbalance twice.

\paragraph*{Rectangular stratification and projections.}

For rectangular treatment grids, we combine
coordinatewise partition refinements through
tensor products and prioritize the resulting
features by refinement depth and projection
dimension, connecting F-OMA to existing
space-filling criteria.

Consider the rectangular treatment grid
$\mathcal X=\{0,\ldots,s^p-1\}^m$,
where $s\ge2$ and $p\ge1$ are integers,
and each coordinate has $s^p$ levels
represented by $p$ base-$s$ digits.
For $a\in\{0,\ldots,s^p-1\}$, let
$f_r(a)=\lfloor a/s^{p-r}\rfloor\bmod s$
denote its $r$th base-$s$ digit, $r=1,\ldots,p$.
Following \citet{tianxu2022}, define 
$\chi_{\mathbf u}(\mathbf x)
=\exp\{(2\pi\mathrm{i}/s)
\sum_{j=1}^m\sum_{r=1}^p
f_{p-r+1}(u_j)f_r(x_j)\}$,
where $\mathbf u,\mathbf x\in\mathcal X$
and $\mathrm{i}^2=-1$.
Define the Niederreiter--Rosenbloom--Tsfasman
(NRT) weight by $\rho(0)=0$ and
$\rho(a)=p+1-\min\{r:f_r(a)\ne0\}$
for $a\ne0$.
For $\mathbf u\in\mathcal X$, let
$\rho(\mathbf u)=\sum_{j=1}^m\rho(u_j)$
and $\pi(\mathbf u)=
\sum_{j=1}^m\mathbb I\{u_j\ne0\}$
denote its total NRT weight and
projection dimension, respectively.
Writing $\chi_{\mathbf u}(d)
=\sum_{i=1}^N\chi_{\mathbf u}(\mathbf x_i)$,
the stratification pattern \citep{tianxu2022}
and projection pattern \citep{shixu2024}
have components
$S_t(d)=N^{-2}\sum_{\rho(\mathbf u)=t}
|\chi_{\mathbf u}(d)|^2$
and
$\Pi_{t,k}(d)=N^{-2}
\sum_{\rho(\mathbf u)=t,\,\pi(\mathbf u)=k}
|\chi_{\mathbf u}(d)|^2$,
respectively.

To construct real treatment-function spaces,
let $\boldsymbol\chi_{\mathbf u}
=(\chi_{\mathbf u}(\mathbf x^{(i)}))_{i=1}^n$
and define
$\mathcal C(I)=\operatorname{span}_{\mathbb R}
\{\operatorname{Re}\boldsymbol\chi_{\mathbf u},
\operatorname{Im}\boldsymbol\chi_{\mathbf u}:
\mathbf u\in I\}\subseteq\mathbb R^n$.
Since both $\rho(\mathbf u)$ and
$\pi(\mathbf u)$ are invariant under
character conjugation, all index sets $I$
used below are closed under conjugation,
and hence $\dim\mathcal C(I)=|I|$.
The Tian--Xu filtration is
$\mathcal H_t^{\mathrm{TX}}
=\mathcal C(\{\mathbf u:\rho(\mathbf u)\le t\})$,
with orthogonal increments
$\mathcal S_t^{\mathrm{TX}}
=\mathcal C(\{\mathbf u:\rho(\mathbf u)=t\})$,
for $t=1,\ldots,mp$.
The Shi--Xu filtration further refines each
layer by projection dimension.
Ordering $(t,k)$ first by $t$ and then by $k$,
define $\mathcal H_{t,k}^{\mathrm{SX}}
=\mathcal C(\{\mathbf u:\rho(\mathbf u)<t
\text{ or }[\rho(\mathbf u)=t,\,
\pi(\mathbf u)\le k]\})$,
with orthogonal increments
$\mathcal S_{t,k}^{\mathrm{SX}}
=\mathcal C(\{\mathbf u:\rho(\mathbf u)=t,\,
\pi(\mathbf u)=k\})$.
Since $k\leq t\leq pk$ is necessary for a nonempty
index set, zero-dimensional increments are omitted.
The following proposition establishes their
connections to F-OMA, with proof given in
Appendix~\ref{app:proof-rectangular}.

\begin{proposition}
\label{prop:rectangular-pattern}
Let $K_t^{\mathrm{TX}}(d;\mathbf V)$ and
$K_{t,k}^{\mathrm{SX}}(d;\mathbf V)$ denote
the F-OMA layer scores corresponding to
the Tian--Xu and Shi--Xu filtrations,
respectively.
With $n=s^{mp}$ and $\mathbf V=\mathbf J_N$,
these scores satisfy
$K_t^{\mathrm{TX}}(d;\mathbf J_N)=S_t(d)/n$
and
$K_{t,k}^{\mathrm{SX}}(d;\mathbf J_N)
=\Pi_{t,k}(d)/n$.
Furthermore, let $\operatorname{SD}_y(d)$ denote
the stratified $L_2$-discrepancy of
\citet{tianxu2026}, using their Theorem 4
weights with $0<y<1$.
Then
$\operatorname{SD}_y(d)^2
=n(1-y)^{-m}\sum_{t=1}^{mp}
y^t K_t^{\mathrm{TX}}(d;\mathbf J_N)$.
\end{proposition}

\paragraph*{Example 2: Rectangular versus
triangular stratification.}

\begin{figure}[tbp]
\centering
\includegraphics[
width=.92\textwidth,
keepaspectratio
]{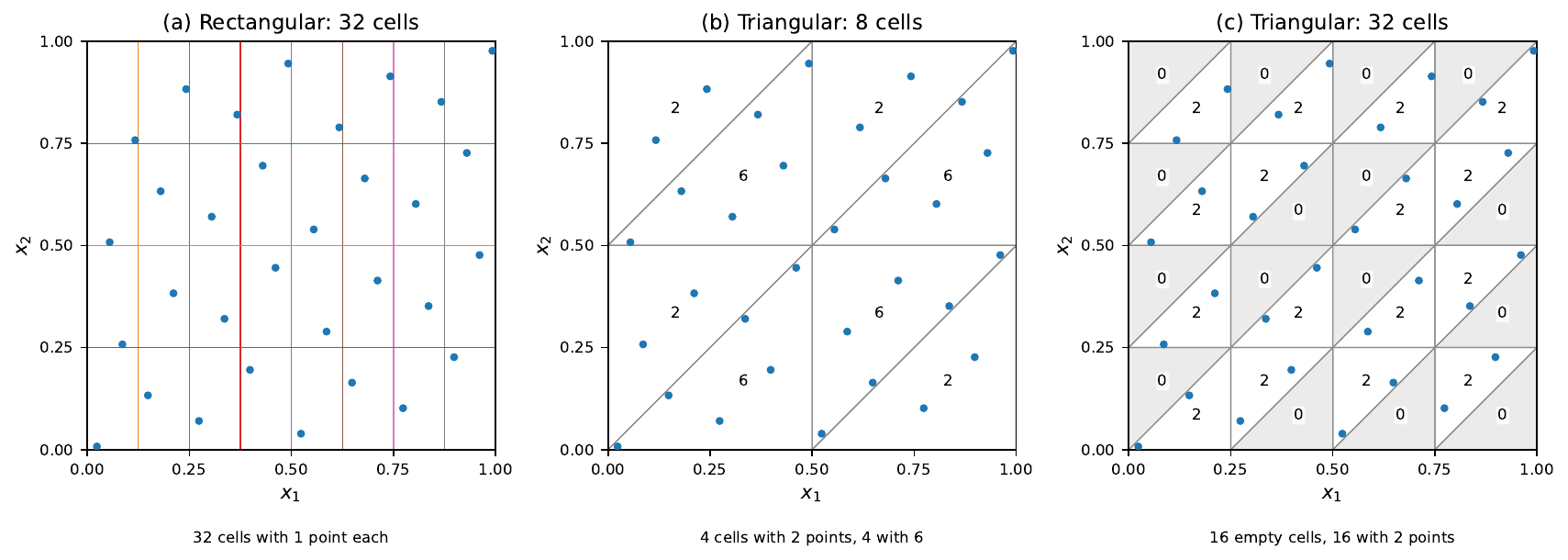}
\caption{Rectangular and triangular stratifications
of the same 32-point design:
(a) $8\times4$ rectangular partition;
(b) triangular partition at depth 2;
(c) triangular partition at depth 3.
Numbers denote cell counts, and shaded
triangles indicate empty cells.}
\label{fig:triangular-rectangular}
\end{figure}

Consider the same treatment region
$\mathcal X=[0,1]^2$ and a 32-point design
with coordinates
$\mathbf x^{(i)}
=((i+3/4)/32,
(\operatorname{rev}_5(i)+1/4)/32)^\top$,
for $i=0,\ldots,31$,
where $\operatorname{rev}_5(i)$ is the integer
obtained by reversing the five binary digits
of $i$, including leading zeros.
For every $a=0,\ldots,5$, the dyadic
rectangular partition into
$2^a\times2^{5-a}$ cells contains
exactly one point per cell.
Thus, the design achieves perfect
rectangular stratification through
total refinement depth five, as in
the Tian--Xu framework.
In contrast, under the triangular
partitions of Example 1, four cells
at depth two contain two points and
four contain six points, giving
$\Delta_2(d;\nu)=1/4$.
At depth three, 16 cells are empty
and 16 contain two points, yielding
$\Delta_3(d;\nu)=1$.
Since $\Delta_1(d;\nu)=1/16$,
Proposition~\ref{prop:partition-foma}
gives the F-OMA score vector
$\mathbf K^\nu(d;\mathbf J_{32})
=(1/16,3/16,3/4)$.
Figure~\ref{fig:triangular-rectangular}
illustrates these differences.
Thus, different treatment
filtrations capture distinct geometric
properties of the same design.

\section{Theoretical Properties and Statistical Justifications}
\label{sec:theory}

Of particular interest is PR-OBS, which
encompasses classical orthogonal block
structures while allowing certain unequal
block sizes \citep{nelder1965,chang2025}.
Using its established stratum decomposition,
we characterize optimality under unknown
stratum weights.
We then provide a variance interpretation
of F-OMA for general positive semidefinite
unit kernels.

\subsection{Stratum Representation under PR-OBS}
\label{subsec:pr-obs-theory}

A PR-OBS is defined by a collection $\mathcal{B}$
of unit factors \citep{chang2025}.
Each unit factor partitions the experimental
units into disjoint, nonempty blocks.
For unit factors $F$ and $G$, write $F\preceq G$
if every block of $F$ is contained in a block
of $G$, and $F\prec G$ if $F\preceq G$ and
$F\ne G$.
Let $U$ and $E$ denote the universal and
singleton factors, respectively.
The join $F\vee G$ is their finest common
coarsening.
Following \citet{chang2025}, a collection
$\mathcal{B}$ is a PR-OBS if it contains
$U$ and $E$, is closed under joins, and
its unit factors are \emph{pairwise orthogonal}.
Using the projectors $\mathbf P_F$
defined in Section~\ref{subsec:unit-kernels},
the orthogonality condition requires:
\(
\mathbf P_F\mathbf P_G
=\mathbf P_G\mathbf P_F
=\mathbf P_{F\vee G},
\)
for all $F,G\in\mathcal B.$

Starting with the coarser factors, define the
orthogonal stratum projectors recursively as
$\mathbf E_F=\mathbf P_F-
\sum_{G\in\mathcal B:\,F\prec G}\mathbf E_G$.
Some stratum projectors may be zero because the coarser strata
already span $\operatorname{col}(\mathbf P_F)$.
We therefore let
$\mathcal B_+=\{F\in\mathcal B:\mathbf E_F\ne\mathbf0\}$
index the nonzero strata.
Under the PR-OBS conditions, the nonzero
$\mathbf E_F$ are mutually orthogonal projectors
whose sum is $\mathbf I_N$.
The recursion gives
$\mathbf P_G=\sum_{F\in\mathcal B_+:\,G\preceq F}
\mathbf E_F$,
whose inversion is the M\"obius inversion
on the factor poset \citep[Theorem~6.9]{bailey2004}.
Thus, the factor-projector unit kernel
in Section~\ref{subsec:unit-kernels} becomes
\begin{equation}
\label{eq:pr-obs-weight-transform}
\mathbf V
=\sum_{G\in\mathcal B}w_G\mathbf P_G
=\sum_{F\in\mathcal B_+}\xi_F\mathbf E_F,
\qquad
\xi_F=\sum_{\substack{G\in\mathcal B\\G\preceq F}}w_G.
\end{equation}
For $w_G\ge0$, the stratum weights are
nonnegative and satisfy
$F\preceq H\implies\xi_F\le\xi_H$.
Substituting the PR-OBS decomposition
\eqref{eq:pr-obs-weight-transform} into
\eqref{eq:foma-projector} gives
$K_t(d;\mathbf V)
=N^{-2}\sum_{F\in\mathcal B_+}
\xi_F\|\mathbf E_F\mathbf Z_t(d)\|_{\mathrm F}^2$,
where $\|\cdot\|_{\mathrm F}$ denotes the
Frobenius norm.
Each term measures the alignment of layer-$t$
treatment features with a unit stratum.
Consequently, $K_t(d;\mathbf V)=0$ if and
only if $\mathbf E_F\mathbf Z_t(d)=\mathbf0$
for every stratum with $\xi_F>0$.

\paragraph{Optimality under Unknown Stratum Weights}
\label{subsec:robust-optimality}

For a given PR-OBS, the stratum projectors
are fixed, but their weights may be unknown.
We study simultaneous optimality over the
monotone cone
$\Xi_{\mathcal B}
=\{\boldsymbol\xi\in[0,\infty)^{\mathcal B_+}:
F\preceq H\Longrightarrow\xi_F\le\xi_H\}$,
which contains, but need not equal, the set of
stratum-weight vectors induced by nonnegative
factor-projector weights in
\eqref{eq:pr-obs-weight-transform}.
For each $\boldsymbol{\xi}\in\Xi_{\mathcal{B}}$, write
$\mathbf{V}_{\boldsymbol{\xi}}
=\sum_{F\in\mathcal{B}_+}\xi_F\mathbf{E}_F$.
A subset $\mathcal G\subseteq\mathcal B_+$
is an \emph{upper set} if, for any
$F,H\in\mathcal B_+$, the conditions
$F\in\mathcal G$ and $F\preceq H$
imply $H\in\mathcal G$.
Let $\mathfrak U(\mathcal B_+)$ denote
the collection of nonempty upper sets.
For each $\mathcal G\in\mathfrak U(\mathcal B_+)$,
let $\boldsymbol\xi^{\mathcal G}\in\Xi_{\mathcal B}$
be the indicator vector with components
$\xi_F^{\mathcal G}=\mathbb I\{F\in\mathcal G\}$.
We write
$\mathbf V_{\mathcal G}:=\mathbf V_{\boldsymbol\xi^{\mathcal G}}
=\sum_{F\in\mathcal G}\mathbf E_F$
for the corresponding kernel.
The following result extends the unknown-weight
comparison of \citet{chang2025} to general
treatment filtrations.
The proof is given in
Appendix~\ref{app:proof-unknown-stratum-weights}.
\begin{theorem}
\label{thm:unknown-stratum-weights}
Fix a treatment filtration and an admissible
design class $\mathcal{D}$.
A design $d^\star\in\mathcal{D}$ lexicographically
minimizes
$\mathbf{K}(d;\mathbf{V}_{\boldsymbol{\xi}})$
over $\mathcal{D}$ for every
$\boldsymbol{\xi}\in\Xi_{\mathcal{B}}$
if and only if it lexicographically minimizes
$\mathbf{K}(d;\mathbf{V}_{\mathcal{G}})$
over $\mathcal{D}$ for every
$\mathcal{G}\in\mathfrak{U}(\mathcal{B}_+)$.
The same equivalence holds for minimizing
any fixed layer $K_t$.
\end{theorem}

If the binary scenarios share a common minimizer,
that design is F-OMA-optimal for all admissible stratum
weights. Otherwise, no simultaneously optimal
design exists and the weights or a compromise criterion
must be specified.

\subsection{Variance Interpretation of F-OMA}
\label{subsec:variance-interpretation}

Consider the response model
$Y_i=f(\mathbf x_i)+\alpha_i+
\zeta(\mathbf x_i,u_i)+\varepsilon_i$,
where $f$ is a fixed treatment-response
function, $\alpha_i$ is a random unit effect,
$\zeta(\mathbf x_i,u_i)$ is a treatment--unit
interaction, and $\varepsilon_i$ is observational
noise.
Assume all random components have mean zero,
with $\operatorname{Cov}(\boldsymbol\alpha)
=\boldsymbol\Sigma_\alpha$ and
$\operatorname{Cov}(\boldsymbol\varepsilon)
=\sigma_\varepsilon^2\mathbf I_N$.
This formulation is related to mixed-effects models
for blocked experiments \citep{hinkelmannkempthorne2008}.

Experimental-unit heterogeneity may affect
different treatment features to different
degrees.
We model this variation through a
filtration-based random-coefficient expansion,
$\zeta(\mathbf x,u)=
\sum_{t=1}^T\zeta_t(\mathbf x,u)$, where
$\zeta_t(\mathbf x,u)=
\tau_t\sum_{\ell=1}^{r_t}
q_{t,\ell}(\mathbf x)b_{t,\ell}(u)$
and $\tau_t>0$.
Here $q_{t,\ell}$ denotes the treatment
function whose evaluations form the
$\ell$th column of $\mathbf Q_t$.
The mean-zero random coefficients satisfy
$\operatorname{Cov}\{b_{t,\ell}(u_i),
b_{s,k}(u_j)\}
=\delta_{ts}\delta_{\ell k}(\mathbf V)_{ij}$,
where $\delta$ is the Kronecker delta.
Assume further that
$\operatorname{Cov}(\boldsymbol{\alpha},
\boldsymbol{\varepsilon})=\mathbf{0}$
and that each coefficient vector
$\mathbf{b}_{t,\ell}=(b_{t,\ell}(u_i))_{i=1}^{N}$
is uncorrelated with both
$\boldsymbol{\alpha}$ and $\boldsymbol{\varepsilon}$.
Under this model, the treatment--unit
interaction covariance is determined
jointly by the treatment filtration and
the unit kernel.
The following theorem relates the variance
of each interaction layer to its F-OMA score.
The proof is given in
Appendix~\ref{app:proof-foma-variance}.

\begin{theorem}
\label{thm:foma-variance}
For a design $d$, let
$\overline{\zeta}_{t,d}
=N^{-1}\sum_{i=1}^N
\zeta_t(\mathbf x_i,u_i)$.
Then
$\operatorname{Var}(\overline{\zeta}_{t,d})
=\tau_t^2K_t(d;\mathbf V)$.
\end{theorem}

Theorem~\ref{thm:foma-variance} identifies
the contribution of each interaction layer
to the variance of the response average.
Define $\bar Y_d=N^{-1}\sum_{i=1}^N Y_i$
and $\bar f_d=N^{-1}\sum_{i=1}^N f(\mathbf x_i)$.
By Theorem~\ref{thm:foma-variance} and the
assumed cross-component uncorrelatedness,
we obtain
\begin{equation}
\label{eq:response-level-variance}
\operatorname{Var}(\bar Y_d-\bar f_d)
=
\frac{\mathbf1_N^\top
\boldsymbol\Sigma_\alpha\mathbf1_N}{N^2}
+\sum_{t=1}^T\tau_t^2K_t(d;\mathbf V)
+\frac{\sigma_\varepsilon^2}{N}.
\end{equation}
Reducing the variance of sample averages
is an established objective in structured
sampling designs \citep{mckay1979}.
Since $\mathbb E(\bar Y_d)=\bar f_d$,
the variance in
\eqref{eq:response-level-variance}
equals the mean squared error of $\bar Y_d$
for estimating $\bar f_d$.
For fixed covariance parameters, only
$\sum_{t=1}^T\tau_t^2K_t(d;\mathbf V)$
depends on the design.
By Proposition~\ref{prop:foma-scalarization},
choosing $\tau_t^2=\epsilon^{t-1}$
for sufficiently small $\epsilon>0$
makes minimizing this variance equivalent
to lexicographically minimizing
$\mathbf K(d;\mathbf V)$ over a fixed
feasible design class.

Appendix~\ref{app:mmd-results} provides
an MMD
interpretation of F-OMA scores and
associated reproducing kernel Hilbert
space (RKHS) integration bounds.
Additionally, under an estimable treatment
model with a positive definite
residual covariance $\sigma^2\mathbf V$,
the layer-weighted integrated mean squared
prediction error, denoted by $\operatorname{IMSE}_{\mathbf a}(d)$,
satisfies
\[
\operatorname{IMSE}_{\mathbf a}(d)
\le
\frac{\sigma^2}{n\gamma_{\mathbf a}(d)^2}
\left\{
a_0K_0(\mathbf V)
+\sum_{t=1}^{T_0}a_tK_t(d;\mathbf V)
\right\},
\]
where $a_t>0$ weights the constant
($t=0$) and $T_0$ retained layers,
$K_0(\mathbf V)$ is the constant-layer
score, and $\gamma_{\mathbf a}(d)>0$
measures weighted sampled-feature coverage.
For a fixed unit kernel $\mathbf{V}$, a fixed $a_0>0$, and a fixed treatment multiset satisfying
estimability, $\gamma_{\mathbf{a}}(d)$ is invariant under
unit reassignment.
Consequently, Proposition~\ref{prop:foma-scalarization} implies that choosing
$a_t=\epsilon^{t-1}$ for $t=1,\ldots,T_0$ and
sufficiently small $\epsilon>0$ makes any globally
F-OMA-optimal assignment also minimize this upper bound.
Details are given in Appendix~\ref{app:global-prediction}.

Appendix~\ref{app:manifold-robot-arm} illustrates these prediction-error
relationships using a manifold-based filtration
and a non-PR-OBS temporal unit kernel.


\section{Numerical Studies}
\label{sec:numerical-studies}

We compare three initializers for Bayesian optimization (BO)
in synthetic experiments under nested and crossed unit structures:
F-OMA, simple random sampling (SRS), and a maximum-projection
Latin hypercube design mapped to the discrete grid
(MaxProLHD$\to$grid; \citealp{joseph2015}).
Each initializer is paired with unit-aware and
unit-blind Gaussian process (GP) surrogates, giving six methods.
For each structure, we conduct 100 paired replicates with
36 initial evaluations followed by 36 BO steps.

\subsection{Experimental Setup}
\label{subsec:numerical-setup}

\paragraph*{Treatments and units.}
The candidate space is
$\mathcal X=\{0,1,2\}^2\times\{0,1\}^3$, with $n=72$.
The five treatment factors represent hypothetical
hyperparameters: learning rate, batch size, hidden width,
optimizer, and dropout. We use the factorial-order filtration
in Section~\ref{subsec:effect-hierarchy} and assign 36 distinct
configurations to 36 units. The nested structure has three
$F_1$ blocks, each containing three $F_2$ blocks, each of which
contains two $F_3$ blocks of two units. The crossed structure
has four rows and nine columns, with one unit per intersection.
Both are orthogonal block structures and hence satisfy
the PR-OBS conditions in Section~\ref{subsec:pr-obs-theory}.
Using the corresponding stratum projectors,
the F-OMA design kernels are
$\mathbf V_{\mathrm{des}}^{\mathrm{NS}}
=\mathbf P_U+\mathbf E_{F_1}+\mathbf E_{F_2}+\mathbf E_{F_3}
=\mathbf P_{F_3}$ and
$\mathbf V_{\mathrm{des}}^{\mathrm{RC}}
=\mathbf P_U+\mathbf E_R+\mathbf E_C$.

\paragraph*{Initial designs.}
F-OMA uses Algorithm~\ref{alg:foma-exchange}, enforcing distinct
configurations and balanced one-factor margins. SRS samples
36 configurations uniformly without replacement.
MaxProLHD$\to$grid maps a continuous design to the discrete
grid and repairs duplicate configurations. Neither baseline
optimizes treatment--unit alignment. For each structure,
one F-OMA design is fixed across all replicates, whereas
both baselines are regenerated in each replicate.

\paragraph*{Responses.}
In replicate $b$, the loss at evaluation $i$ is
\begin{equation}
\label{eq:simulation-response}
Y_{b,i}=R(\mathbf x_{b,i})+g_b(u_{b,i})
+h_b(\mathbf x_{b,i},u_{b,i})+\varepsilon_{b,i},
\end{equation}
where $R$ is a fixed treatment surface, $g_b$ is a unit effect,
$h_b$ is a treatment--unit interaction, and $\varepsilon_{b,i}$
is independent observation noise within each trajectory.
Both $g_b$ and $h_b(\mathbf x,\cdot)$ are centered over units,
so $R$ is the environment-averaged expected loss.
The two structures share $R$ and the noise level.
Response generation and unit-aware GP fitting use a kernel
$\mathbf{K}_u$ distinct from $\mathbf{V}_{\mathrm{des}}$.
Appendix~\ref{app:simulation-details} describes the response generator, initialization
procedures, GP fitting, and sequential optimization.

\subsection{Bayesian Optimization and Performance Measures}
\label{subsec:bo-performance}

\paragraph*{Surrogates and sequential evaluations.}
Both GPs use an unknown constant mean and a treatment
correlation combining Mat\'ern-$5/2$ for ordered factors
with a categorical correlation for optimizer.
The unit-aware GP adds unit effects and treatment--unit
interactions using the simulator's coefficient-kernel
matrix $\mathbf K_u$; the blind GP omits these terms.
The remaining GP parameters are fitted separately
from each method's observations, as detailed in Appendix~\ref{app:simulation-details}.
Within each replicate, all methods share latent effects,
common random numbers, and a response-independent random
permutation of the 36 follow-up units. The two GPs for each
initializer receive identical initial observations.
At each step, the next unit is known, and knowledge gradient
\citep{frazier2009} selects the treatment with the largest
expected decrease in the minimum posterior mean of the
environment-averaged loss. All 72 candidates remain eligible,
including previously evaluated treatments.

\paragraph*{Performance measures.}
Let $\widehat R_{b,m,t}(\mathbf x)$ denote the fitted
environment-averaged loss for replicate $b$, method $m$,
and BO step $t$, where $t=0$ follows the initial batch.
The recommendation is
$\widehat{\mathbf{x}}_{b,m,t}\in
\operatorname*{arg\,min}_{\mathbf{x}\in\mathcal X}
\widehat R_{b,m,t}(\mathbf{x})$,
with regret
$r_{b,m}(t)=R(\widehat{\mathbf{x}}_{b,m,t})-R^\star$,
where $R^\star=\min_{\mathbf{x}\in\mathcal X}R(\mathbf{x})$.
Recommendations are selected from the full candidate set,
including treatments absent from the observations;
recommendation regret need not decrease monotonically. 
The true $R$ is used only for
evaluation. The normalized area under the recommendation-regret
curve (AURC) is
\begin{equation}
\label{eq:bo-aurc}
A_{b,m}(H)=\frac{1}{2H}\sum_{t=0}^{H-1}
\{r_{b,m}(t)+r_{b,m}(t+1)\}.
\end{equation}
For the main horizon $H=36$, we report the mean paired
difference $\bar D=100^{-1}\sum_{b=1}^{100}D_b$,
where $D_b=A_{b,m}(36)-A_{b,m'}(36)$, and unadjusted
95\% paired $t$ intervals over 100 replicates.
Negative differences favor method $m$.
These trajectory-level intervals quantify Monte Carlo
uncertainty conditional on the fixed F-OMA designs and
treatment surface. Results for $H=10,18$ and best-evaluated
target loss are in Appendix~\ref{app:additional-bo-results}.

\subsection{Results and Discussion}
\label{subsec:numerical-results}

\begin{figure}[p]
\centering
\renewcommand{\baselinestretch}{1}
\normalsize
\setlength{\abovecaptionskip}{4pt}
\setlength{\belowcaptionskip}{0pt}

\begin{minipage}[t]{0.49\textwidth}
\centering
\includegraphics[width=\linewidth]
  {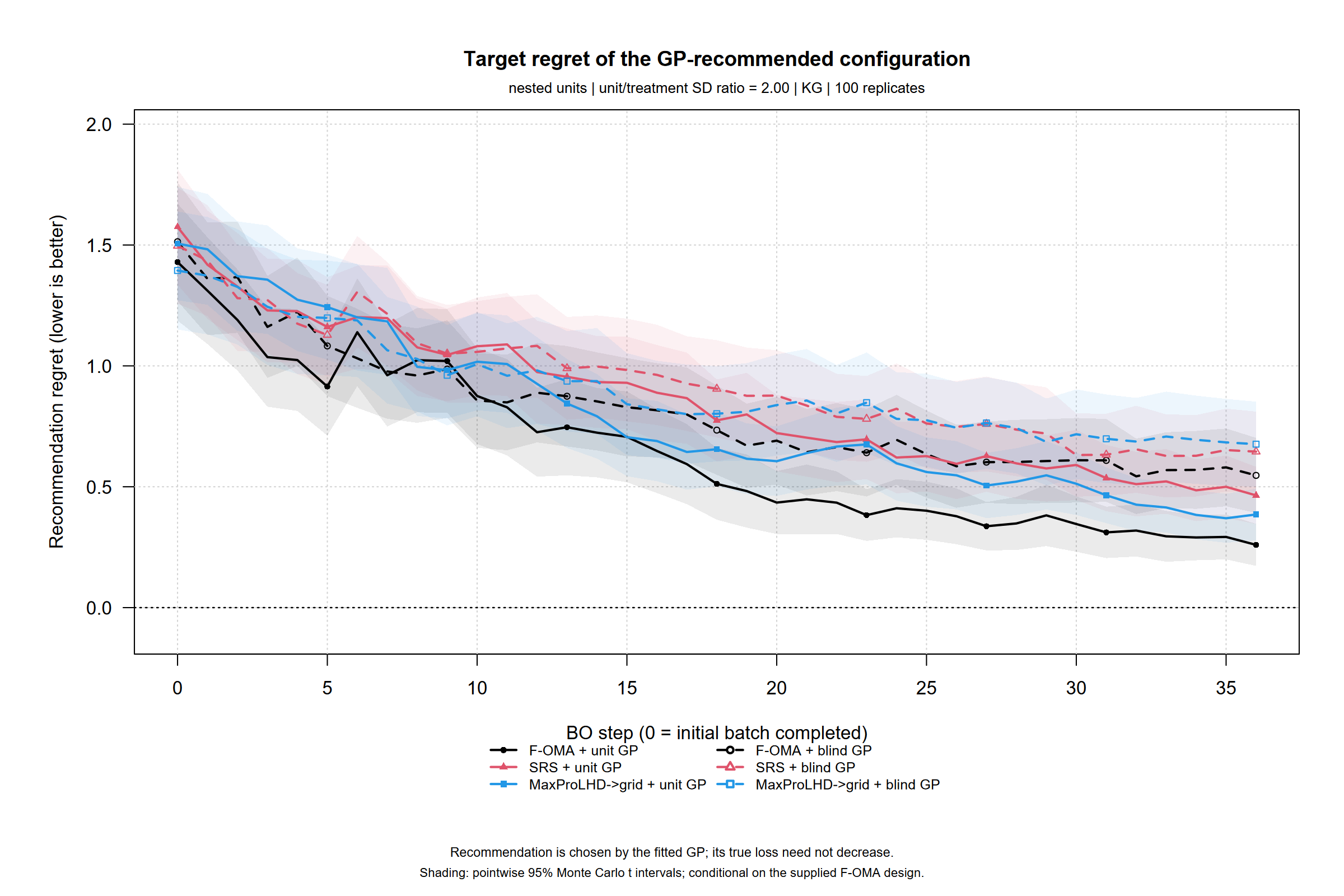}
\par\vspace{2pt}
{\small (a) Nested units\par}
\end{minipage}\hfill
\begin{minipage}[t]{0.49\textwidth}
\centering
\includegraphics[width=\linewidth]
  {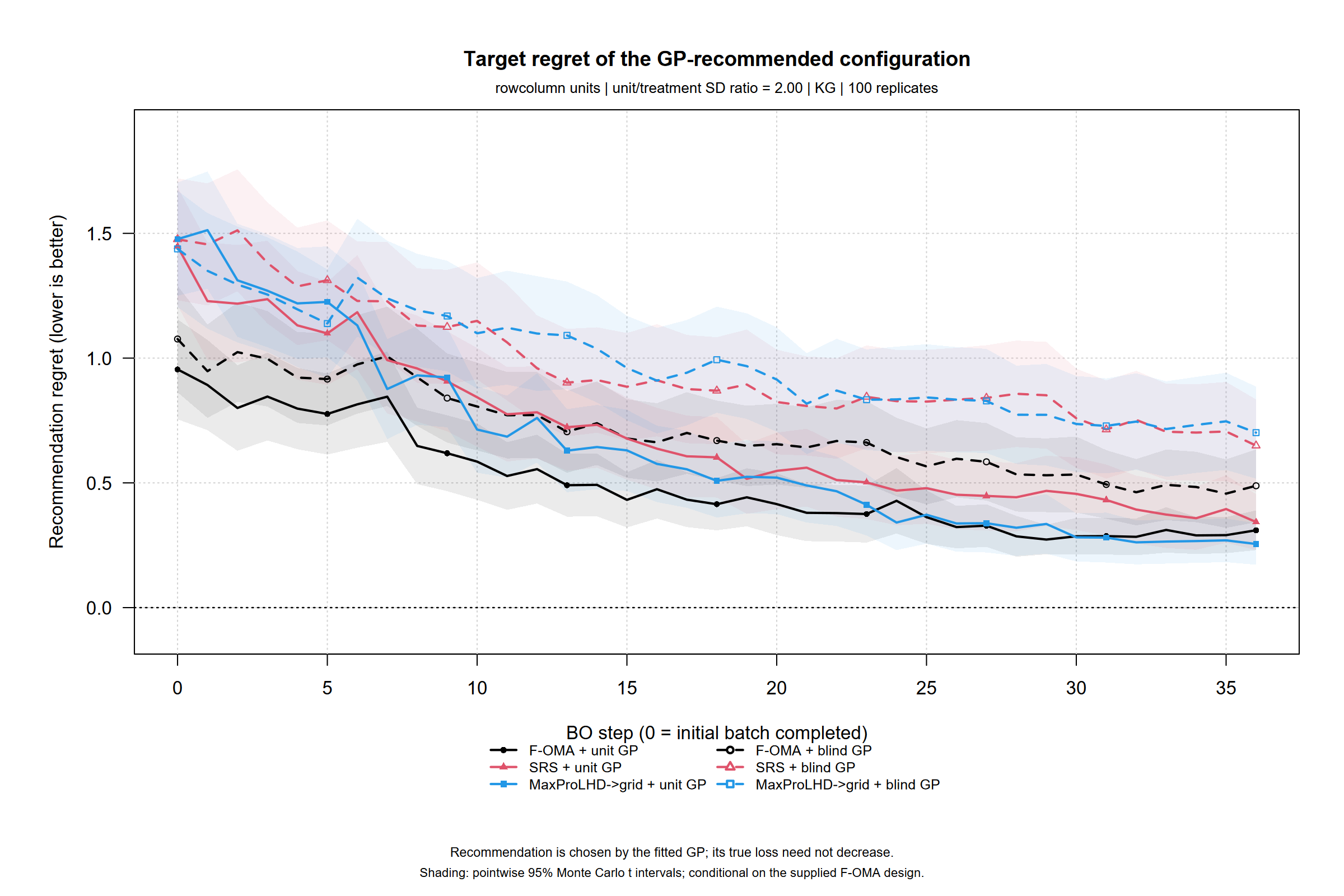}
\par\vspace{2pt}
{\small (b) Crossed units\par}
\end{minipage}

\caption{Mean recommendation regret over 100 paired replicates,
with pointwise 95\% Monte Carlo intervals for each method.
Step~0 follows 36 initial evaluations; each subsequent step adds
one evaluation. Appendix~\ref{app:simulation-details} defines
the RMS ratio in the panel subtitles.}
\label{fig:bo-regret}

\vspace{10pt}

\begin{minipage}[t]{0.49\textwidth}
\centering
\includegraphics[width=\linewidth]
  {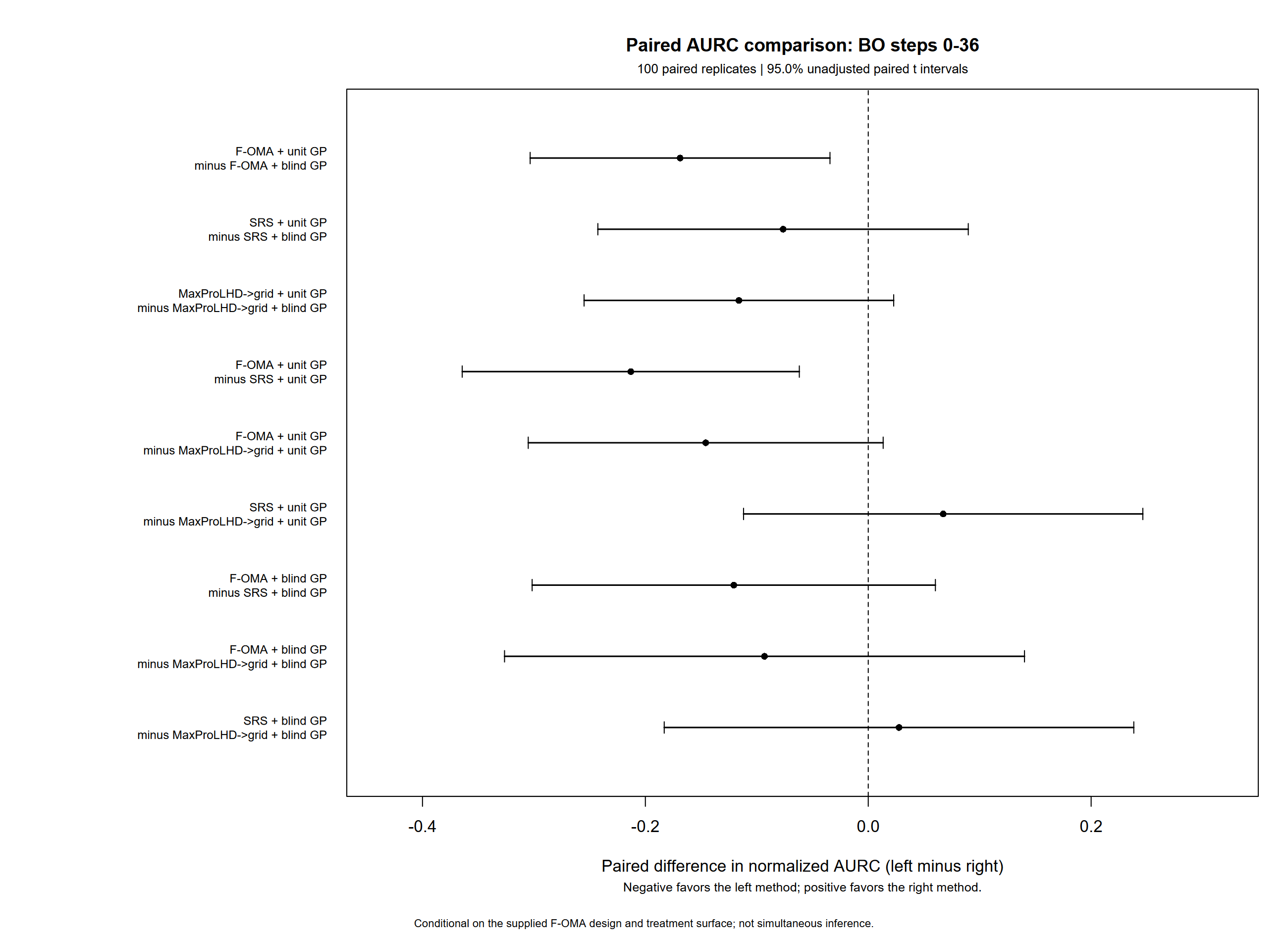}
\par\vspace{2pt}
{\small (a) Nested units\par}
\end{minipage}\hfill
\begin{minipage}[t]{0.49\textwidth}
\centering
\includegraphics[width=\linewidth]
  {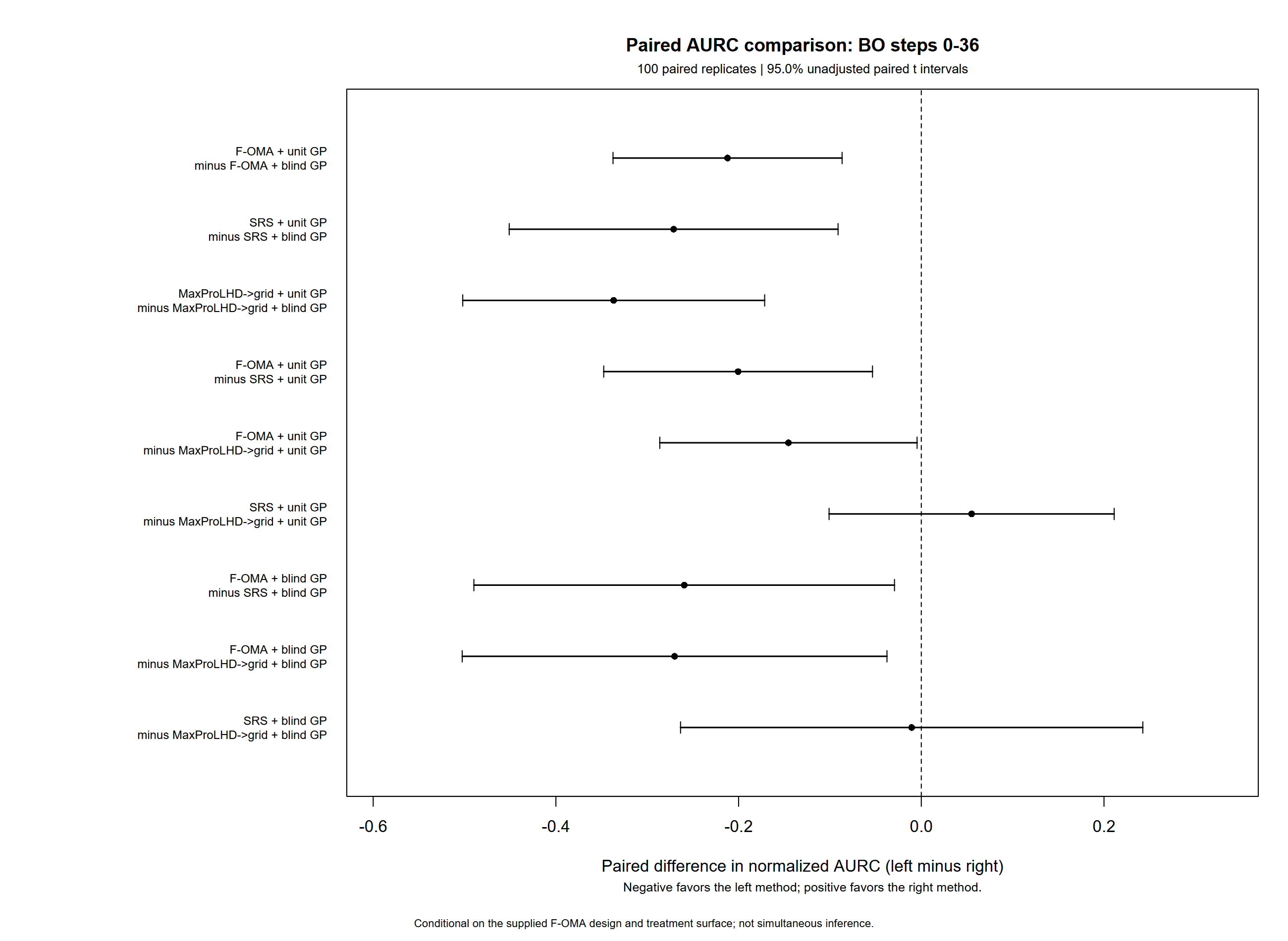}
\par\vspace{2pt}
{\small (b) Crossed units\par}
\end{minipage}

\caption{Mean paired differences in normalized AURC over BO steps~0--36
(first method minus second), with unadjusted 95\% paired $t$
intervals from 100 replicates. Negative values favor the first method.
Intervals in both figures are conditional on the fixed F-OMA
designs and treatment surface.}
\label{fig:bo-aurc-full}
\end{figure}

Figures~\ref{fig:bo-regret} and~\ref{fig:bo-aurc-full}
show recommendation regret and paired full-horizon AURC
differences, respectively.

\paragraph*{Nested units.}
F-OMA with the unit-aware GP has the lowest mean regret
over much of the middle and late follow-up.
Its AURC is lower than that of F-OMA with the blind GP
and SRS with the unit-aware GP, with both paired intervals
below zero. It also has lower mean AURC than
MaxProLHD$\to$grid with the unit-aware GP, but that
interval includes zero.

\paragraph*{Crossed units.}
F-OMA with the unit-aware GP has the lowest mean regret
over much of the early and middle follow-up, whereas
MaxProLHD$\to$grid with the unit-aware GP has slightly
lower final-step mean regret. The unit-aware GP improves
AURC over the blind GP for all three initializers,
with paired intervals below zero. Under the unit-aware GP,
the F-OMA-minus-SRS interval is below zero; the
F-OMA-minus-MaxProLHD$\to$grid estimate is negative,
with its upper interval endpoint close to zero.
Thus, final-step regret and trajectory-average regret
give different comparisons.

These findings support F-OMA initialization with unit-aware
modeling in the examined settings, without uniform superiority.
The best-evaluated-loss results in
Appendix~\ref{app:additional-bo-results} further distinguish
sampling a good configuration from correctly recommending it
using heterogeneous observations.


\clearpage

\FloatBarrier

\FloatBarrier
\section{Discussion}
\label{sec:discussion}

F-OMA provides a model-free framework accommodating
effect-hierarchy, geometric, manifold-based, spectral,
and multiresolution treatment filtrations, together
with independently specified positive semidefinite
unit kernels encompassing PR-OBS and beyond.
It recovers classical aberration and certain space-filling
criteria, characterizes simultaneous optimality under
unknown monotone PR-OBS stratum weights, and provides
variance and MMD interpretations, a GLS prediction-error
bound under additional assumptions, and a constrained
multistart algorithm.

The numerical studies illustrate the potential benefits of
combining F-OMA initialization with unit-aware surrogate modeling.
Under both nested and crossed unit structures, F-OMA yields
lower mean AURC than SRS when both use a unit-aware GP over
$H=36$ BO steps (Section~\ref{sec:numerical-studies}), although the gains vary across
comparators and evaluation horizons.
These findings highlight the complementary roles of initial
design construction and surrogate modeling in BO with
heterogeneous experimental units.
Natural extensions of F-OMA include data-adaptive treatment
filtrations, robust design under general unit-kernel uncertainty,
and computationally efficient sequential design augmentation.

\section*{Software Availability}

The \texttt{FOMA} R package, together with its documentation
and illustrative examples, is available
\href{https://www.dropbox.com/scl/fo/u2xl6o0zejfzptp9hw9l4/AGEqHr-yQtuUFFOAtH9WhZo?rlkey=hmqyrj56bn9l4n6tql34d1jtj&dl=0}{here}.

\bibliographystyle{agsm}
\bibliography{references}

\newpage

\appendix

\section{Proofs of Main Results}
\label{app:section3}

\subsection{Proof of Proposition~\ref{prop:foma-scalarization}}
\label{app:proof-foma-scalarization}

\begin{proof}
Fix the treatment filtration and unit kernel,
and set $a_t=\epsilon^{t-1}$.
For two designs $d,d'\in\mathcal{D}$ with
different score vectors, write
$\Delta_t=K_t(d;\mathbf{V})-K_t(d';\mathbf{V})$,
and let $m$ be the first index with $\Delta_m\ne0$.
The difference between their scalar scores is
$\epsilon^{m-1}
\{\Delta_m+\sum_{t=m+1}^T
\epsilon^{t-m}\Delta_t\}$.
The expression in braces converges to
$\Delta_m\ne0$ as $\epsilon\downarrow0$.
Thus, the scalar-score difference has the same
sign as $\Delta_m$ for all sufficiently small
positive $\epsilon$.
Since $\mathcal{D}$ is finite, there are only
finitely many such pairs.
Taking the minimum of their positive thresholds
gives a common $\epsilon_0>0$.
Designs with identical score vectors have
identical scalar scores for every $\epsilon>0$.
If all score vectors coincide, take $\epsilon_0=1$.
Therefore, for every $0<\epsilon<\epsilon_0$,
$A_{\mathbf{a}}(d;\mathbf{V})$ preserves the
lexicographic ordering, including ties.
\end{proof}

\subsection{Proof of Proposition~\ref{prop:classical-aberration}}
\label{app:classical-aberration}

\begin{proof}
Recall that $\mathbf{r}_d=\mathbf{X}_d^\top\mathbf{1}_N$.
For $\mathbf{V}=\mathbf{J}_N$, the trace representation
gives $K_t(d;\mathbf{J}_N)
=N^{-2}\|\mathbf{Q}_t^\top\mathbf{r}_d\|_2^2$.
For each factor $j$, choose real-valued functions
$c_{j,0},\ldots,c_{j,s_j-1}$ satisfying
$c_{j,0}=1$ and
$\sum_{x_j\in\mathcal{X}_j}
c_{j,a}(x_j)c_{j,b}(x_j)=s_j\delta_{ab}$,
where $\delta_{ab}$ is the Kronecker delta.
For $\boldsymbol{\alpha}=(\alpha_1,\ldots,\alpha_m)$
with $0\leq\alpha_j\leq s_j-1$, define
$c_{\boldsymbol{\alpha}}(\mathbf{x})
=\prod_{j=1}^m c_{j,\alpha_j}(x_j)$
and
$\mathbf{q}_{\boldsymbol{\alpha}}
=n^{-1/2}
(c_{\boldsymbol{\alpha}}(\mathbf{x}))_{\mathbf{x}\in\mathcal{X}}$.
Tensor-product orthogonality implies that these
vectors form an orthonormal basis of $\mathbb{R}^n$.
Let
$\mu_{\boldsymbol{\alpha}}(d)
=N^{-1}\sum_{i=1}^N
c_{\boldsymbol{\alpha}}(\mathbf{x}_i)$.
Then
$\mathbf{q}_{\boldsymbol{\alpha}}^\top\mathbf{r}_d
=Nn^{-1/2}\mu_{\boldsymbol{\alpha}}(d)$.

\textit{Factorial-order filtration.}
Let $\operatorname{wt}(\boldsymbol{\alpha})
=\sum_{j=1}^m\mathbb{I}\{\alpha_j>0\}$.
The vectors $\mathbf{q}_{\boldsymbol{\alpha}}$
with $\operatorname{wt}(\boldsymbol{\alpha})=t$
form an orthonormal basis of
$\mathcal{S}_t^{\mathrm{fac}}$.
By basis invariance,
$K_t^{\mathrm{fac}}(d;\mathbf{J}_N)
=n^{-1}\sum_{\operatorname{wt}(\boldsymbol{\alpha})=t}
\mu_{\boldsymbol{\alpha}}(d)^2$.
Under the stated normalization, the sum is
the generalized wordlength component
$A_t(d)$ of \citet{xuwu2001}.
Hence $K_t^{\mathrm{fac}}(d;\mathbf{J}_N)=A_t(d)/n$.

\textit{Polynomial-degree filtration.}
For quantitative factors, take
$c_{j,a}=\sqrt{s_j}\rho_{j,a}$,
using the polynomial basis and numerical levels
specified in Section~\ref{subsec:effect-hierarchy}.
Then $\mathbf{q}_{\boldsymbol{\alpha}}$ is the
evaluation vector of $\rho_{\boldsymbol{\alpha}}$.
The vectors with
$|\boldsymbol{\alpha}|=\sum_{j=1}^m\alpha_j=t$
form an orthonormal basis of
$\mathcal{S}_t^{\mathrm{poly}}$.
Consequently,
$K_t^{\mathrm{poly}}(d;\mathbf{J}_N)
=n^{-1}\sum_{|\boldsymbol{\alpha}|=t}
\mu_{\boldsymbol{\alpha}}(d)^2$.
In the indicator-function representation of
\citet{chengye2004}, the coefficients satisfy
$b_{\boldsymbol{\alpha}}
=n^{-1}\sum_{i=1}^N
c_{\boldsymbol{\alpha}}(\mathbf{x}_i)$
and $b_{\mathbf{0}}=N/n$.
Thus,
$b_{\boldsymbol{\alpha}}/b_{\mathbf{0}}
=\mu_{\boldsymbol{\alpha}}(d)$,
so the sum equals $\beta_t(d)$.
Hence $K_t^{\mathrm{poly}}(d;\mathbf{J}_N)=\beta_t(d)/n$.

\end{proof}

\subsection{Proof of Proposition~\ref{prop:partition-foma}}
\label{app:proof-partition}

\begin{proof}
For each $s=0,\ldots,T$, the projection
kernel satisfies
\[
\sum_{i,j=1}^N
P_s^\nu(\mathbf x_i,\mathbf x_j)V_{ij}
=\sum_{C\in\Pi_s}
\mathbf e_C(d)^\top
\mathbf V\mathbf e_C(d)/\nu(C).
\]
Using $p_t^\nu=P_t^\nu-P_{t-1}^\nu$
in the definition of $K_t^\nu$
proves \eqref{eq:partition-general}.
For $\mathbf V=\mathbf J_N$, we have
$\mathbf e_C(d)^\top\mathbf J_N
\mathbf e_C(d)=N_C(d)^2$.
Since $\sum_{C\in\Pi_s}N_C(d)=N$
and $\sum_{C\in\Pi_s}\nu(C)=1$,
expanding \eqref{eq:partition-discrepancy}
gives
$\Delta_s(d;\nu)
=N^{-2}\sum_{C\in\Pi_s}
N_C(d)^2/\nu(C)-1$.
Taking the difference between levels
$t$ and $t-1$ in
\eqref{eq:partition-general}
proves \eqref{eq:partition-foma-j}.
Finally, for a finite uniform reference
measure, $\nu(C)=n_C/n$, so
\eqref{eq:partition-general} reduces to
$K_t^\nu(d;\mathbf V)=nK_t(d;\mathbf V)$,
recovering the finite-space formulation.
\end{proof}

\subsection{Proof of Proposition~\ref{prop:rectangular-pattern}}
\label{app:proof-rectangular}

\begin{proof}
Let $n=s^{mp}$ and define
$\mathbf{q}_{\mathbf{u}}
=n^{-1/2}
(\chi_{\mathbf{u}}(\mathbf{x}))_{\mathbf{x}\in\mathcal{X}}$
for $\mathbf{u}\in\mathcal{X}$, using the
inverse-digit characters of \citet{tianxu2022}.
Character orthogonality implies that these
vectors form an orthonormal basis of $\mathbb{C}^n$.
Throughout this proof, ${}^*$ denotes
the conjugate transpose.

For one factor, characters of NRT weight
at most $r$ depend only on the first $r$
most significant base-$s$ digits.
There are $s^r$ such orthogonal characters,
equal to the dimension of this function space.
Thus, characters of weight exactly $r$ span
the orthogonal increment from depth $r-1$
to depth $r$.
Tensor products give the groups indexed by
total NRT weight $\rho(\mathbf{u})$.
Further grouping by $\pi(\mathbf{u})$
specifies the number of nonconstant
coordinate factors.
Complex conjugation corresponds to digitwise
negation modulo $s$, which preserves both
$\rho(\mathbf{u})$ and $\pi(\mathbf{u})$.
Hence both types of character groups are
closed under conjugation.
Their projectors are therefore real symmetric
and admit real orthonormal bases, consistent
with the real-valued F-OMA formulation.
For either group $\mathcal{I}$, its projector is
$\mathbf{B}_{\mathcal{I}}
=\sum_{\mathbf{u}\in\mathcal{I}}
\mathbf{q}_{\mathbf{u}}\mathbf{q}_{\mathbf{u}}^*$.
Recall that
$\mathbf{r}_d=\mathbf{X}_d^\top\mathbf{1}_N$
and
$\chi_{\mathbf{u}}(d)
=\sum_{i=1}^N\chi_{\mathbf{u}}(\mathbf{x}_i)$.
Then
$\mathbf{q}_{\mathbf{u}}^*\mathbf{r}_d
=n^{-1/2}\overline{\chi_{\mathbf{u}}(d)}$.
The F-OMA score is therefore
$K_{\mathcal{I}}(d;\mathbf{J}_N)
=N^{-2}\sum_{\mathbf{u}\in\mathcal{I}}
|\mathbf{q}_{\mathbf{u}}^*\mathbf{r}_d|^2
=(nN^2)^{-1}\sum_{\mathbf{u}\in\mathcal{I}}
|\chi_{\mathbf{u}}(d)|^2$.
Taking
$\mathcal{I}=\{\mathbf{u}:\rho(\mathbf{u})=t\}$
gives
$K_t^{\mathrm{TX}}(d;\mathbf{J}_N)=S_t(d)/n$
by the definition of \citet{tianxu2022}.
Taking
$\mathcal{I}
=\{\mathbf{u}:\rho(\mathbf{u})=t,\,
\pi(\mathbf{u})=k\}$
gives
$K_{t,k}^{\mathrm{SX}}(d;\mathbf{J}_N)
=\Pi_{t,k}(d)/n$
by the definition of \citet{shixu2024}.
This proves both identities.

Finally, \citet[Theorem~4]{tianxu2026}
show that, under their specified
stratification weights,
$\operatorname{SD}_y(d)^2
=(1-y)^{-m}\{E(d;y)-1\}$,
where
$E(d;y)=1+\sum_{t=1}^{mp}S_t(d)y^t$
is the stratification pattern enumerator.
Since $S_t(d)=nK_t^{\mathrm{TX}}(d;\mathbf{J}_N)$,
substitution gives
$\operatorname{SD}_y(d)^2
=n(1-y)^{-m}\sum_{t=1}^{mp}
y^tK_t^{\mathrm{TX}}(d;\mathbf{J}_N)$,
as claimed.

\end{proof}

\subsection{Proof of Theorem~\ref{thm:unknown-stratum-weights}}
\label{app:proof-unknown-stratum-weights}

We first give a finite representation of the
monotone weight cone.

\begin{lemma}[Upper-set representation]
\label{lem:upper-set-representation}
Every vector in $\Xi_{\mathcal{B}}$ is a
nonnegative linear combination of the vectors
$\boldsymbol{\xi}^{\mathcal{G}}$,
$\mathcal{G}\in\mathfrak{U}(\mathcal{B}_+)$.
Conversely, every such combination belongs
to $\Xi_{\mathcal{B}}$.
\end{lemma}

\begin{proof}
For $\boldsymbol{\xi}=\mathbf{0}$, use the empty sum.
Otherwise, since $\mathcal{B}_+$ is finite,
let $0<\alpha_1<\cdots<\alpha_L$
be the distinct positive coordinate values
of $\boldsymbol{\xi}$, and set $\alpha_0=0$.
Define
$\mathcal{G}_\ell
=\{F\in\mathcal{B}_+:\xi_F\geq\alpha_\ell\}$
for $\ell=1,\ldots,L$.
Monotonicity of $\boldsymbol{\xi}$ makes each
$\mathcal{G}_\ell$ a nonempty upper set.
Coordinatewise,
$\boldsymbol{\xi}
=\sum_{\ell=1}^L
(\alpha_\ell-\alpha_{\ell-1})
\boldsymbol{\xi}^{\mathcal{G}_\ell}$.
Conversely, each upper-set indicator satisfies
the defining inequalities of $\Xi_{\mathcal{B}}$,
as does any nonnegative linear combination.
\end{proof}

\begin{proof}[Proof of Theorem~\ref{thm:unknown-stratum-weights}]
Necessity follows because every binary
upper-set vector belongs to $\Xi_{\mathcal{B}}$.

For sufficiency, suppose that the same design
$d^\star$ lexicographically minimizes every
binary scenario.
Fix $d\in\mathcal{D}$ and
$\boldsymbol{\xi}\in\Xi_{\mathcal{B}}$.
If $\boldsymbol{\xi}=\mathbf{0}$, all scores vanish.
Otherwise, Lemma~\ref{lem:upper-set-representation}
gives
$\boldsymbol{\xi}
=\sum_{\ell=1}^L b_\ell
\boldsymbol{\xi}^{\mathcal{G}_\ell}$
with $b_\ell>0$.
Write
$\boldsymbol{\delta}_\ell
=\mathbf{K}(d;\mathbf{V}_{\mathcal{G}_\ell})
-\mathbf{K}(d^\star;\mathbf{V}_{\mathcal{G}_\ell})$.
Linearity of the stratum representation gives
$\mathbf{K}(d;\mathbf{V}_{\boldsymbol{\xi}})
-\mathbf{K}(d^\star;\mathbf{V}_{\boldsymbol{\xi}})
=\sum_{\ell=1}^L b_\ell\boldsymbol{\delta}_\ell$.
Each $\boldsymbol{\delta}_\ell$ is zero or has
a positive first nonzero coordinate.
If all are zero, the two designs have identical
scores under $\mathbf{V}_{\boldsymbol{\xi}}$.
Otherwise, let $t_0$ be the earliest nonzero
coordinate among these vectors.
All preceding coordinates vanish, and every
$t_0$th coordinate is nonnegative, with at least
one strictly positive.
Since $b_\ell>0$, their weighted sum is
lexicographically positive.
As $d$ and $\boldsymbol{\xi}$ were arbitrary,
$d^\star$ is optimal throughout $\Xi_{\mathcal{B}}$.
For the fixed-layer statement, assume instead
that $d^\star$ minimizes $K_t$ under every
binary scenario.
Then every $t$th component of
$\boldsymbol{\delta}_\ell$ is nonnegative,
so the same linearity identity gives
$K_t(d;\mathbf{V}_{\boldsymbol{\xi}})
-K_t(d^\star;\mathbf{V}_{\boldsymbol{\xi}})\geq0$.
Together with necessity, this proves the
fixed-layer equivalence.
\end{proof}

\begin{corollary}
\label{cor:omit-identity-scenario}
For a fixed run size $N$, the all-strata scenario
can be omitted from
Theorem~\ref{thm:unknown-stratum-weights}
if its score vector is constant over $\mathcal{D}$.
This holds if every layer kernel has a constant
diagonal on $\mathcal{X}$, or if all designs
in $\mathcal{D}$ have the same treatment
replication vector.
In particular, if
$p_t(\mathbf{x},\mathbf{x})=c_t$
for every $\mathbf{x}\in\mathcal{X}$,
then $K_t(d;\mathbf{I}_N)=c_t/N$.
\end{corollary}

\begin{proof}
By the PR-OBS stratum decomposition,
$\mathbf{V}_{\mathcal{B}_+}
=\sum_{F\in\mathcal{B}_+}\mathbf{E}_F
=\mathbf{I}_N$.
The kernel definition gives
$K_t(d;\mathbf{I}_N)
=N^{-2}\sum_{i=1}^N
p_t(\mathbf{x}_i,\mathbf{x}_i)$.
If $p_t$ has constant diagonal $c_t$,
this equals $c_t/N$.
Alternatively, writing $r_{d,j}$ for the
$j$th entry of $\mathbf{r}_d$ gives
$K_t(d;\mathbf{I}_N)
=N^{-2}\sum_{j=1}^n
r_{d,j}p_t(\mathbf{x}^{(j)},\mathbf{x}^{(j)})$.
Thus, a fixed replication vector also makes
every layer score constant over $\mathcal{D}$.
In either case, every feasible design minimizes
the all-strata score vector, so this scenario
adds no restriction to the simultaneous
optimality conditions.
\end{proof}

\subsection{Proof of Theorem~\ref{thm:foma-variance}}
\label{app:proof-foma-variance}

\begin{proof}
For any $\mathbf V\succeq\mathbf0$,
let $\mathbf g_{t,\ell}$ be independent
random vectors with mean zero and covariance
$\mathbf I_N$.
Setting
$\mathbf b_{t,\ell}
=(b_{t,\ell}(u_i))_{i=1}^N
=\mathbf V^{1/2}\mathbf g_{t,\ell}$
gives
$\operatorname{Cov}
(\mathbf b_{t,\ell},\mathbf b_{s,k})
=\delta_{ts}\delta_{\ell k}\mathbf V$.
Thus, the assumed coefficient model exists
for any positive semidefinite unit kernel.
By the definition of the treatment kernel,
$p_t(\mathbf x,\mathbf x')
=\sum_{\ell=1}^{r_t}
q_{t,\ell}(\mathbf x)q_{t,\ell}(\mathbf x')$.
The coefficient covariance assumption gives
$\operatorname{Cov}
\{\zeta_t(\mathbf x,u_i),
\zeta_s(\mathbf x',u_j)\}
=\delta_{ts}\tau_t^2
p_t(\mathbf x,\mathbf x')(\mathbf V)_{ij}$.
Consequently,
$\operatorname{Var}(\overline{\zeta}_{t,d})
=\tau_t^2N^{-2}\sum_{i,j=1}^N
p_t(\mathbf x_i,\mathbf x_j)(\mathbf V)_{ij}
=\tau_t^2K_t(d;\mathbf V)$,
proving the theorem.
The result requires only the stated
first two moments, not independence
or Gaussianity.
\end{proof}

\section{Additional Treatment Filtrations and Illustrations}
\label{app:filtration-illustrations}

\subsection{Unequal Allocation under
Nested Rectangular Partitions}
\label{app:partition-allocation}

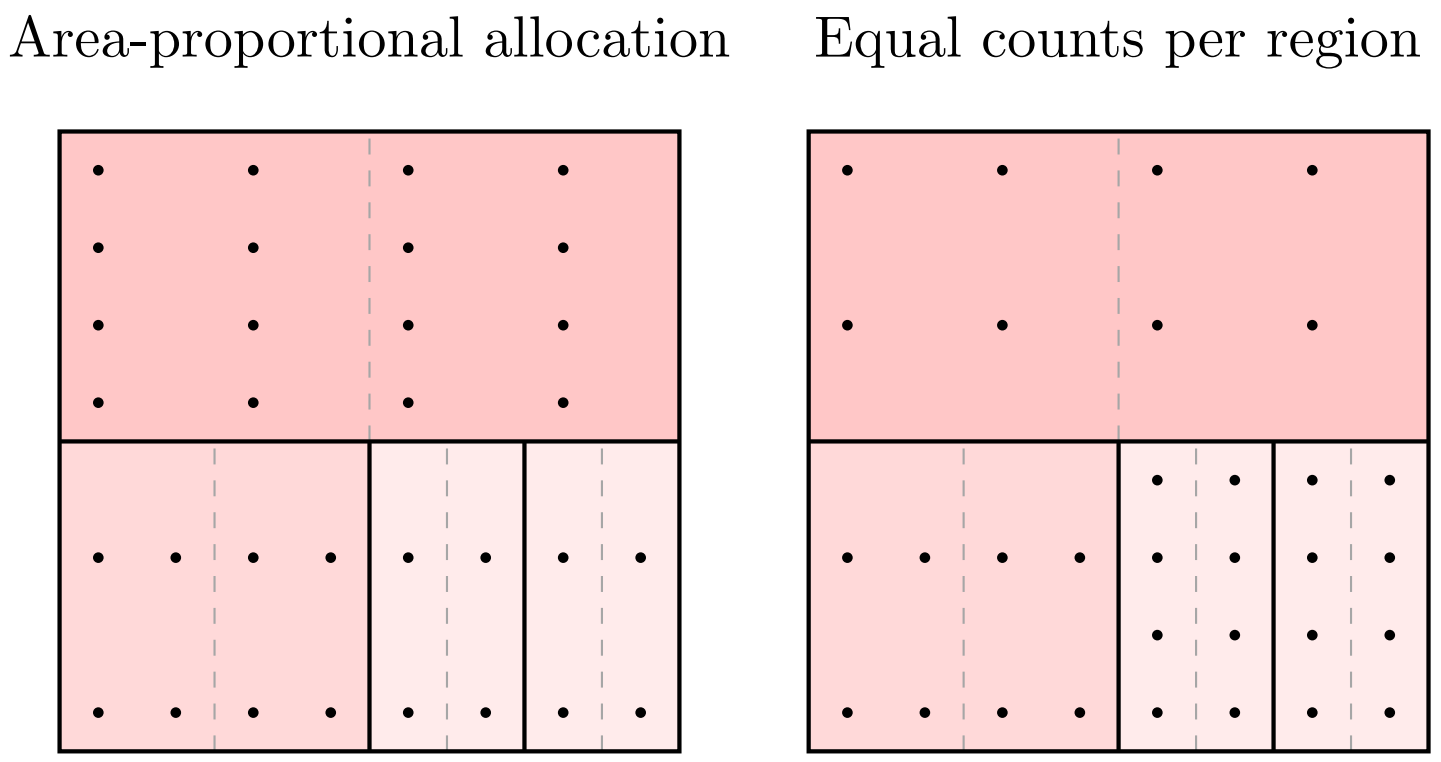
\begin{figure}[t]
\centering
\begingroup
\newcommand{\DrawAllocation}[1]{%
\begin{tikzpicture}[x=4.1cm,y=4.1cm]

\fill[red!22] (0,0.5) rectangle (1,1);
\fill[red!15] (0,0) rectangle (0.5,0.5);
\fill[red!8] (0.5,0) rectangle (0.75,0.5);
\fill[red!8] (0.75,0) rectangle (1,0.5);

\draw[gray!70,thin,dashed]
  (0.5,0.5)--(0.5,1);
\draw[gray!70,thin,dashed]
  (0.25,0)--(0.25,0.5);
\draw[gray!70,thin,dashed]
  (0.625,0)--(0.625,0.5);
\draw[gray!70,thin,dashed]
  (0.875,0)--(0.875,0.5);

\draw[black,thick] (0,0.5)--(1,0.5);
\draw[black,thick] (0.5,0)--(0.5,0.5);
\draw[black,thick] (0.75,0)--(0.75,0.5);
\draw[black,thick] (0,0) rectangle (1,1);

\ifnum#1=1

  \foreach \i in {0,2,4,6}{
    \foreach \j in {4,5,6,7}{
      \fill ({(\i+0.5)/8},{(\j+0.5)/8})
        circle[radius=1pt];
    }
  }
  \foreach \i in {0,1,2,3,4,5,6,7}{
    \foreach \j in {0,2}{
      \fill ({(\i+0.5)/8},{(\j+0.5)/8})
        circle[radius=1pt];
    }
  }

\else

  \foreach \i in {0,2,4,6}{
    \foreach \j in {5,7}{
      \fill ({(\i+0.5)/8},{(\j+0.5)/8})
        circle[radius=1pt];
    }
  }
  \foreach \i in {0,1,2,3}{
    \foreach \j in {0,2}{
      \fill ({(\i+0.5)/8},{(\j+0.5)/8})
        circle[radius=1pt];
    }
  }
  \foreach \i in {4,5,6,7}{
    \foreach \j in {0,1,2,3}{
      \fill ({(\i+0.5)/8},{(\j+0.5)/8})
        circle[radius=1pt];
    }
  }

\fi
\end{tikzpicture}%
}

\begin{tabular}{@{}c@{\hspace{0.5cm}}c@{}}
\small Area-proportional allocation &
\small Equal counts per region\\[5pt]
\DrawAllocation{1} & \DrawAllocation{2}
\end{tabular}
\endgroup

\caption{Allocation under unequal region sizes:
(left) area-proportional allocation and
(right) equal counts per region.}
\label{fig:partition-allocation}
\end{figure}

Figure~\ref{fig:partition-allocation} compares
two designs with $N=32$ distinct points selected
from an $8\times8$ treatment grid ($n=64$).
Under the uniform reference measure $\nu$,
the four coarse regions have probabilities
$(1/2,1/4,1/8,1/8)$, giving target
allocations $(16,8,4,4)$ for $N=32$.
Each region is further divided into two equal-sized
cells, forming a nested partition.
Solid and dashed lines indicate coarse and
refined boundaries, respectively.
The first increment space $\mathcal S_1$
captures contrasts among the four coarse regions,
with $\dim(\mathcal S_1)=4-1=3$.
The second increment space $\mathcal S_2$
captures contrasts between the two refined cells
within each coarse region, with
$\dim(\mathcal S_2)=8-4=4$.

Let $d_{\mathrm{prop}}$ and $d_{\mathrm{eq}}$
denote the left and right designs, respectively.
By \eqref{eq:partition-discrepancy},
$\Delta_1(d_{\mathrm{prop}};\nu)
=\Delta_2(d_{\mathrm{prop}};\nu)=0$,
whereas
$\Delta_1(d_{\mathrm{eq}};\nu)
=\Delta_2(d_{\mathrm{eq}};\nu)=3/8$.
Proposition~\ref{prop:partition-foma}
then gives
$(K_1^\nu,K_2^\nu)=(0,0)$
and $(3/8,0)$, respectively, under
$\mathbf V=\mathbf J_{32}$.
Thus, the equal-count design has coarse-level
imbalance but no additional imbalance from
refinement, illustrating how orthogonal
increments avoid counting the same imbalance twice.

\subsection{Hexagonal Partitions}
\label{app:hexagonal-illustration}

Consider a treatment domain consisting of
32 equal-area hexagonal cells, grouped into
eight disjoint supercells, each containing
four hexagons.
The supercells form the coarse partition
$\Pi_1$, while the individual hexagons form
the fine partition $\Pi_2$.
Each supercell is a fixed union of fine
hexagons and need not be a regular hexagon.
This grouping ensures nestedness, which
independently generated hexagonal grids
need not satisfy.

Let $\nu$ be the normalized area measure
on the hexagonal treatment domain.
Let $\mathcal H_1,\mathcal H_2\subset L^2(\nu)$
be the spaces of functions constant within
the coarse and fine cells, respectively.
With $\mathcal H_0=\operatorname{span}\{1\}$,
the increment spaces
$\mathcal S_r=\mathcal H_r\ominus\mathcal H_{r-1}$
have dimensions
$\dim(\mathcal S_1)=8-1=7$ and
$\dim(\mathcal S_2)=32-8=24$,
representing contrasts among supercells
and within supercells, respectively.

\begin{figure}[p]
\centering
\includegraphics[
    width=\textwidth,
    height=.70\textheight,
    keepaspectratio
]{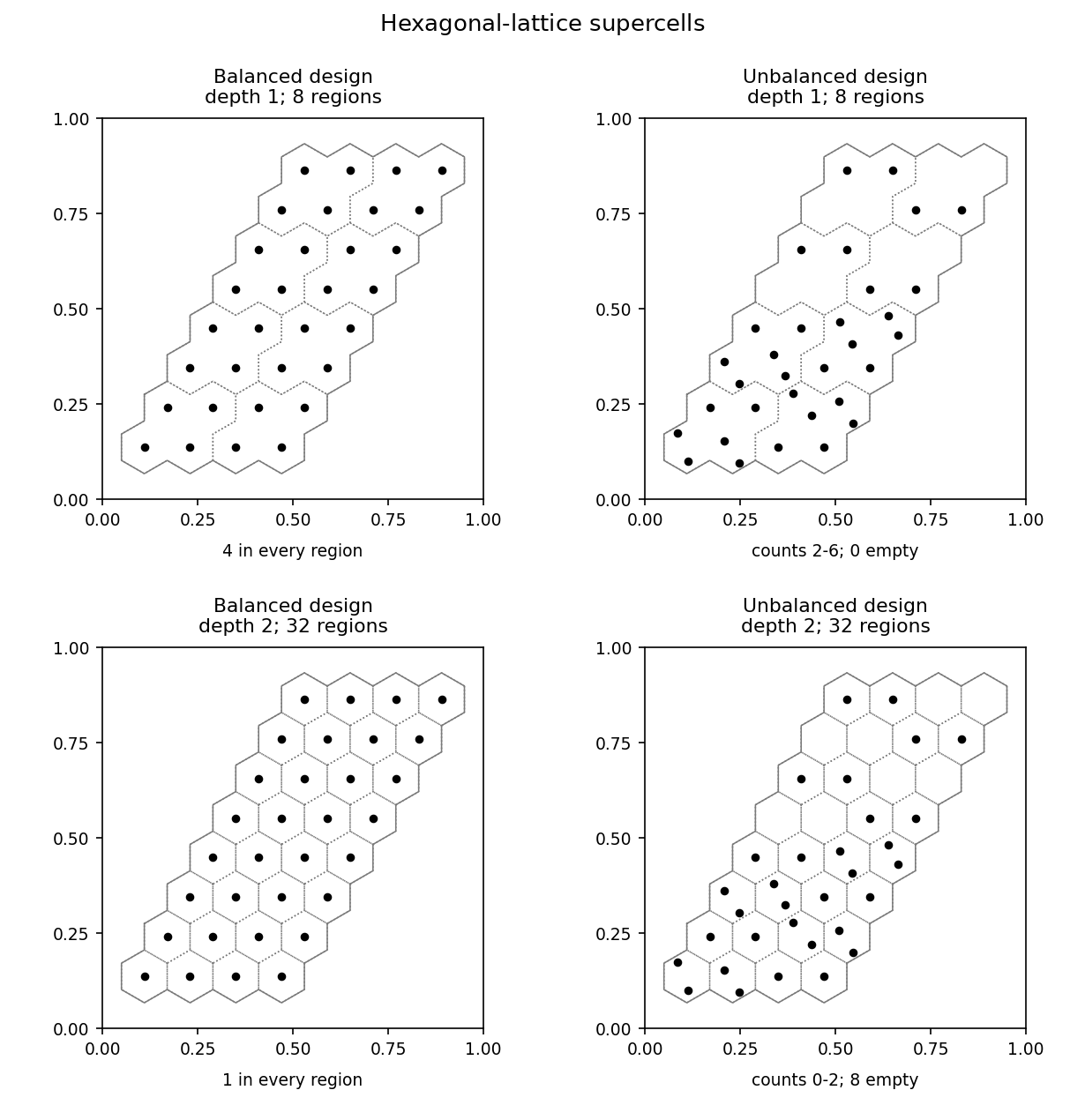}
\caption{Nested hexagonal partitions with
eight coarse supercells and 32 fine cells,
comparing balanced and unbalanced
32-point designs.}
\label{fig:hexagonal}
\end{figure}

Figure~\ref{fig:hexagonal} compares two
32-point designs.
The balanced design places four points
in each supercell and one in each hexagon,
giving zero discrepancy at both levels.
The unbalanced design has eight empty
fine cells, eight cells with two points,
and 16 cells with one point, yielding
$\Delta_2(d;\nu)=1/2$.
The corresponding layer score
$K_2^\nu(d;\mathbf J_N)
=\Delta_2(d;\nu)-\Delta_1(d;\nu)$
measures the additional imbalance at
the finer resolution.
The treatment domain consists only of the
hexagonal cells, not the surrounding
rectangular plotting region.

\subsection{Nested Voronoi Partitions}
\label{app:voronoi-illustration}

Voronoi partitions provide flexible treatment
regions with irregular boundaries.
To obtain a nested filtration, each parent cell
is refined separately rather than by constructing
a new global Voronoi diagram.
For a bounded convex polygon $C$, choose a
direction $\mathbf v$ and a threshold that
divides $C$ into two equal-area regions.
This bisection can be realized by a
parent-restricted Voronoi split using two
nearby interior sites whose perpendicular
bisector gives the dividing line.
Repeating this construction within each parent
produces nested equal-area partitions with
$2^r$ cells at depth $r$.

Let $\nu$ be the normalized area measure
on the treatment domain.
Let $\mathcal H_r\subset L^2(\nu)$ be the
space of functions constant on the cells of
$\Pi_r$, and define
$\mathcal S_r=\mathcal H_r\ominus\mathcal H_{r-1}$,
with $\mathcal H_0=\operatorname{span}\{1\}$.
Since $\Pi_r$ contains $2^r$ cells,
$\dim(\mathcal H_r)=2^r$ and
$\dim(\mathcal S_r)=2^{r-1}$ for $r\ge1$.
In particular, the displayed partitions at
depths 3 and 5 contain 8 and 32 cells,
and the corresponding increment spaces have
dimensions $\dim(\mathcal S_3)=4$,
$\dim(\mathcal S_4)=8$, and
$\dim(\mathcal S_5)=16$.

\begin{figure}[p]
\centering
\includegraphics[
    width=\textwidth,
    height=.73\textheight,
    keepaspectratio
]{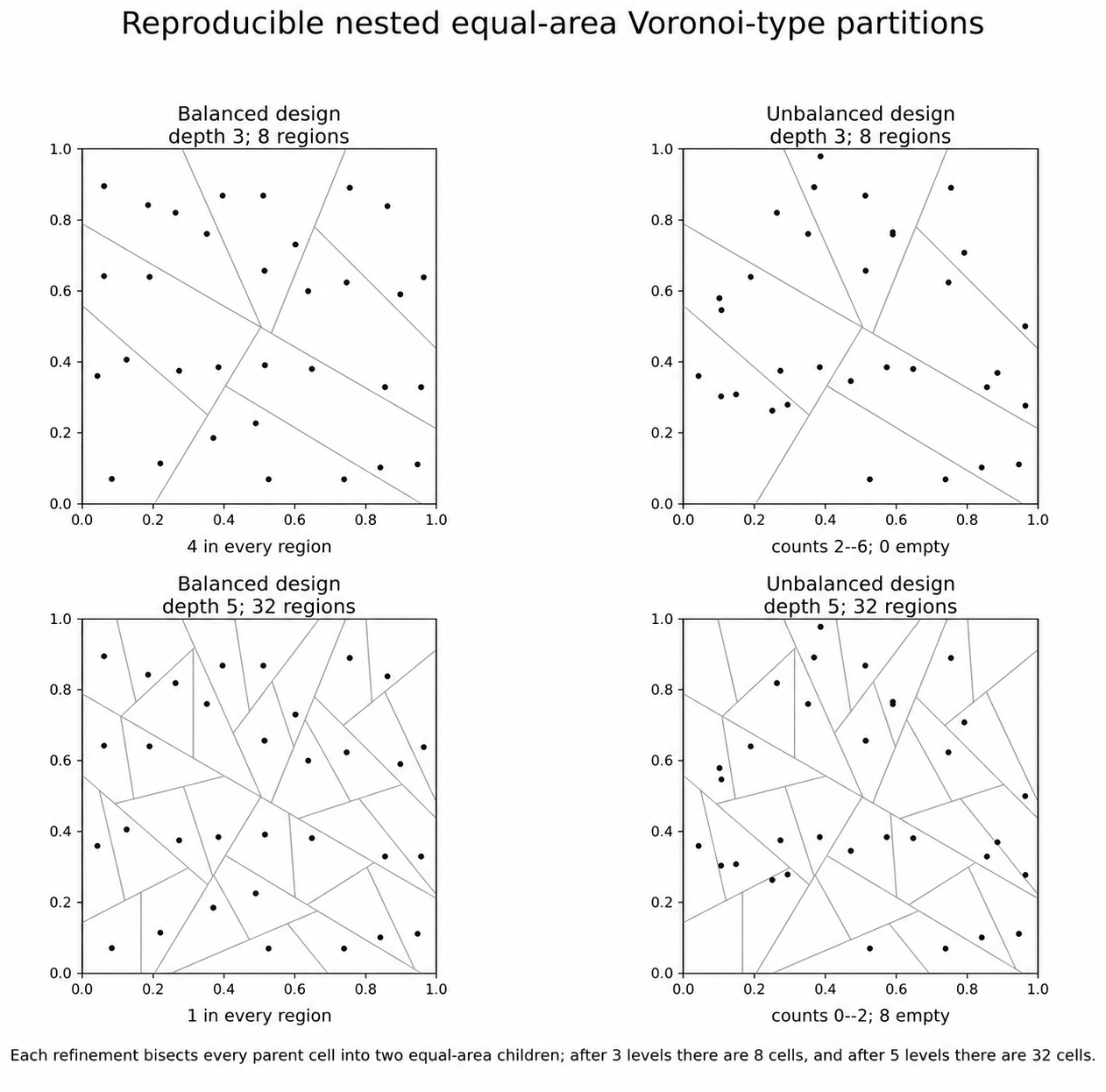}
\caption{Nested equal-area Voronoi-type
partitions at depths 3 and 5, comparing
balanced and unbalanced 32-point designs.}
\label{fig:voronoi}
\end{figure}

Figure~\ref{fig:voronoi} compares two
32-point designs at depths 3 and 5.
The balanced design has four points in each
depth-3 cell and one point in each depth-5 cell,
so its discrepancies are zero at both depths.
For the unbalanced design, the depth-3 cell
counts are $(5,5,2,5,2,6,4,3)$, giving
$\Delta_3(d;\nu)=1/8$.
At depth 5, eight cells are empty, eight
contain two points, and 16 contain one point,
giving $\Delta_5(d;\nu)=1/2$.
Thus, the additional discrepancy revealed
between depths 3 and 5 is
$\Delta_5(d;\nu)-\Delta_3(d;\nu)=3/8$,
which equals the combined contribution of the
intermediate layers.
This example shows that nested refinement can
reveal additional imbalance even when the
treatment regions have irregular polygonal
boundaries.

\FloatBarrier
\subsection{A Manifold-Based Example: A Planar Robot Arm}
\label{app:manifold-robot-arm}

Consider a planar two-link robot arm
\citep{lynchpark2017}, with link lengths
$l_1=1.25$ and $l_2=0.75$ and joint
angles $q_1$ and $q_2$.
Its end-effector position is
$\mathbf x=(l_1\cos q_1+l_2\cos(q_1+q_2),
l_1\sin q_1+l_2\sin(q_1+q_2))^\top$.
With unrestricted joint rotations, the reachable
workspace is the annulus
$\mathcal W=\{\mathbf x\in\mathbb R^2:
0.5\le\|\mathbf x\|_2\le2\}$,
a two-dimensional manifold with boundary.

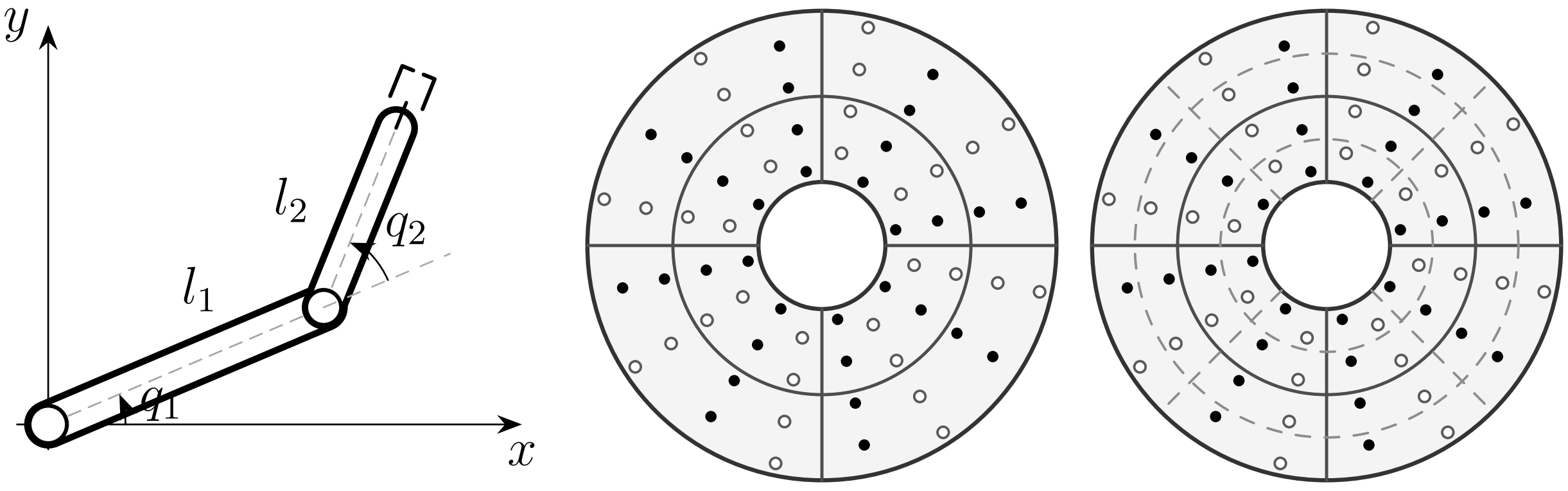
\begin{figure}[t]
\centering

\def\annuluspanel#1{%
\begin{tikzpicture}[line cap=round,line join=round]

  \fill[gray!9] (0,0) circle (1.70);
  \fill[white] (0,0) circle (0.46);

  \draw[black!80,line width=0.85pt]
    (0,0) circle (1.70);
  \draw[black!80,line width=0.85pt]
    (0,0) circle (0.46);
  \draw[black!70,line width=0.65pt]
    (0,0) circle (1.08);

  \foreach \a in {0,90,180,270} {
    \draw[black!70,line width=0.65pt]
      (\a:0.46)--(\a:1.70);
  }

  \ifnum#1=1
    \foreach \r in {0.77,1.39} {
      \draw[black!45,dashed,line width=0.5pt]
        (0,0) circle (\r);
    }
    \foreach \a in {45,135,225,315} {
      \draw[black!45,dashed,line width=0.5pt]
        (\a:0.46)--(\a:1.70);
    }
  \fi

  \foreach \j in {0,1,2,3} {
    \foreach \k in {0,1,2,3,4,5,6,7} {

      \pgfmathsetmacro{\ra}
        {0.46+0.31*(\j+0.28)}
      \pgfmathsetmacro{\rb}
        {0.46+0.31*(\j+0.72)}
      \pgfmathsetmacro{\aa}{45*\k+12}
      \pgfmathsetmacro{\ab}{45*\k+33}

      \fill[black]
        (\aa:\ra) circle (0.041);

      \draw[draw=black!65,fill=white,
            line width=0.5pt]
        (\ab:\rb) circle (0.041);
    }
  }

\end{tikzpicture}%
}

\begin{minipage}[b]{0.36\linewidth}
\centering
\resizebox{\linewidth}{!}{%
\begin{tikzpicture}[
  line cap=round,
  line join=round,
  >={Stealth[length=2mm]}
]

  \def\qone{23}
  \def\qsum{68}

  \coordinate (O) at (0,0);
  \coordinate (E) at (\qone:2.4);
  \coordinate (T) at ($(E)+(\qsum:1.55)$);

  \draw[->,thin]
    (-0.25,0)--(3.8,0)
    node[below] {$x$};
  \draw[->,thin]
    (0,-0.20)--(0,3.20)
    node[left] {$y$};

  \draw[black,line width=11pt]
    (O)--(E);
  \draw[white,line width=8pt]
    (O)--(E);
  \draw[gray!70,densely dashed,thin]
    (O)--(E);

  \draw[black,line width=10pt]
    (E)--(T);
  \draw[white,line width=7pt]
    (E)--(T);
  \draw[gray!70,densely dashed,thin]
    (E)--(T);

  \draw[fill=white,line width=0.9pt]
    (O) circle (0.15);
  \draw[fill=white,line width=0.9pt]
    (E) circle (0.14);

  \draw[->,thin]
    (0.62,0)
    arc[start angle=0,
        end angle=\qone,
        radius=0.62];
  \node at (10:0.92) {$q_1$};

  \draw[gray!65,dashed]
    (E)--($(E)+(\qone:1.1)$);

  \draw[->,thin]
    ($(E)+(\qone:0.56)$)
    arc[start angle=\qone,
        end angle=\qsum,
        radius=0.56];

  \node at ($(E)+(43:0.91)$)
    {$q_2$};

  \node[above=7pt]
    at ($(O)!0.55!(E)$) {$l_1$};
  \node[left=9pt]
    at ($(E)!0.60!(T)$) {$l_2$};

  \begin{scope}[shift={(T)},rotate=\qsum]
    \draw[line width=0.8pt]
      (0,0)--(0.22,0);
    \draw[line width=0.9pt]
      (0.22,-0.14)--(0.48,-0.14)
      --(0.48,-0.045);
    \draw[line width=0.9pt]
      (0.22,0.14)--(0.48,0.14)
      --(0.48,0.045);
  \end{scope}

\end{tikzpicture}%
}

\vspace{2mm}

{\small (a) Two-link robot arm}
\end{minipage}
\hfill
\begin{minipage}[b]{0.30\linewidth}
\centering

\resizebox{\linewidth}{!}{%
  \annuluspanel{0}%
}

\vspace{2mm}

{\small (b) Coarse partition}
\end{minipage}
\hfill
\begin{minipage}[b]{0.30\linewidth}
\centering

\resizebox{\linewidth}{!}{%
  \annuluspanel{1}%
}

\vspace{2mm}

{\small (c) Refined partition}
\end{minipage}

\caption{A planar two-link robot arm and nested
partitions of its reachable workspace:
(a) robot configuration;
(b) coarse partition $\Pi_1$;
(c) refined partition $\Pi_2$.}
\label{fig:manifold-robot-arm}

\end{figure}

Figure~\ref{fig:manifold-robot-arm} illustrates
the robot arm and its reachable workspace.
Panel (a) shows the link lengths $l_1,l_2$
and joint angles $q_1,q_2$, while panels
(b) and (c) show the end-effector positions
in Cartesian coordinates $(x,y)$, with the
robot base at the origin.
Let $\mathcal X\subset\mathcal W$ contain
64 candidate positions.
Dividing the workspace into radial bands and
angular sectors gives eight coarse cells in
$\Pi_1$, each refined into four cells in $\Pi_2$.
Solid and dashed lines indicate coarse and
refined boundaries, respectively.
This defines a nested partition filtration
without requiring a rectangular grid.

Each refined cell contains two candidates.
Under the uniform measure on the
64 candidate positions, selecting one
candidate per refined cell (filled circles)
gives $K_1(d;\mathbf J_{32})
=K_2(d;\mathbf J_{32})=0$.
When trials exhibit temporal dependence,
represented by $V_{ij}=\exp(-\lambda|i-j|)$
with $\lambda>0$, different assignments of
the same treatments may yield different
F-OMA scores.

\paragraph*{Manifold-based design under temporal dependence.}
We extend the planar robot-arm example by assigning
one candidate from each of the 32 fine cells to 32 ordered units.
The treatment set is fixed, and the unit kernel is
$V_{ij}=0.5^{|i-j|}$, which lies outside the PR-OBS stratum-weight
class because its row sums are nonconstant.
We compare random assignment, randomized assignment balanced across
four consecutive eight-unit blocks, and F-OMA assignment-swap search.
Random assignments are uniform permutations of the fixed treatment set.
The block-based baseline assigns one treatment from each coarse cell
to each of four consecutive blocks of eight units, randomizing both
the allocation within each coarse cell across blocks and the order
within blocks. Each baseline is averaged over 10,000 assignments.
F-OMA uses assignment-only search with exhaustive best-improvement
pairwise swaps until no swap improves $(K_1,K_2)$ lexicographically,
starting from the first 100 random assignments.
All starts attain the same scores and IMSEs.
The terminal assignments repeat the same permutation
of the eight coarse cells across the four blocks
and are therefore also block-balanced.
Prediction errors are evaluated under the correctly specified model
of Appendix~\ref{app:global-prediction}, with $n=64$, $N=32$, $T_0=1$, $\sigma^2=1$,
and $a_0=a_1=1$.
The mean response belongs to $\mathcal H_1$, and $\mathbf V$
is treated as the known residual covariance.
Using Euclidean-orthonormal treatment bases, we compute IMSE
analytically for each assignment and integrate uniformly over
the 64 candidates.

\begin{center}
\begin{tabular}{lrrrr}
\hline
Assignment & $10^3K_1$ & $10^3K_2$ & GLS IMSE & \shortstack{\strut OLS IMSE\\\strut (= GLS bound)} \\
\hline
Random & 3.213 & 11.009 & 0.22985 & 0.29545 \\
Block-based & 2.685 & 11.536 & 0.21706 & 0.26170 \\
F-OMA & 2.525 & 11.696 & 0.24750 & 0.25147 \\
\hline
\end{tabular}
\end{center}
Compared with the baseline mean OLS IMSEs, F-OMA gives reductions
of $14.9\%$ relative to random assignment and $3.9\%$ relative to
block-based assignment, with unchanged geometric coverage.
Its GLS IMSE, however, exceeds both baseline means.

The different rankings under OLS and GLS
can be explained by their covariance structures.
Using the notation of Appendix~\ref{app:global-prediction}, let
$\mathbf Z_R=\mathbf X_d[\mathbf q_0,\mathbf Q_1]$,
where $\mathbf q_0=64^{-1/2}\mathbf 1_{64}$, and set
$\mathbf C=\mathbf Z_R^\top\mathbf V\mathbf Z_R$.
The fixed treatment set gives
$\mathbf G=\mathbf Z_R^\top\mathbf Z_R=\mathbf I_8/2$.
With equal prediction weights and $\sigma^2=1$,
\[
\begin{aligned}
\operatorname{IMSE}_{\mathrm{OLS}}(d)
=n^{-1}\operatorname{tr}
(\mathbf G^{-1}\mathbf C\mathbf G^{-1})
=64\{K_0(\mathbf V)+K_1(d;\mathbf V)\},
\end{aligned}
\]
which coincides with the GLS prediction-error bound.
Thus, minimizing the first-layer F-OMA score
in this example minimizes the actual OLS IMSE.
In contrast, GLS exploits the residual covariance
through $\mathbf M=\mathbf Z_R^\top
\mathbf V^{-1}\mathbf Z_R$.
Its improvement over OLS is
\[
\frac{1}{n}\operatorname{tr}
\left(
\mathbf G^{-1}\mathbf C\mathbf G^{-1}
-\mathbf M^{-1}
\right),
\]
which is nonnegative but design-dependent.
The mean reductions from OLS to GLS for random and block-based
assignments are approximately $0.06560$ and $0.04465$, respectively,
whereas the reduction for the reported F-OMA assignment is $0.00397$.
The larger GLS gains for the two baselines
more than offset the OLS advantage of F-OMA.
This illustrates that minimizing a GLS
prediction-error bound need not minimize
the actual GLS prediction error.

\subsection{Spectral and Multiresolution Filtrations}
\label{app:spectral-filtrations}
\label{app:graph-wavelet}

\paragraph*{Graph-spectral filtration.}
Consider a connected undirected graph
with nonnegative edge weights on $\mathcal X$.
Let $\mathbf W_G$ be its adjacency matrix
and $\mathbf L_G=\operatorname{diag}
(\mathbf W_G\mathbf1_n)-\mathbf W_G$
the graph Laplacian.
Let $\mathbf L_G\mathbf u_\ell
=\lambda_\ell\mathbf u_\ell$, where
$0=\lambda_0<\lambda_1\le\cdots
\le\lambda_{n-1}$ and the eigenvectors
are orthonormal, with
$\mathbf u_0=n^{-1/2}\mathbf1_n$.
Choose cutoffs
$0=c_0<c_1<\cdots<c_T=\lambda_{n-1}$
such that each interval $(c_{t-1},c_t]$
contains at least one eigenvalue.
Define
$\mathcal H_t^G
=\operatorname{span}\{\mathbf u_\ell:
\lambda_\ell\le c_t\}$,
with increment projectors
$\mathbf B_t^G
=\sum_{c_{t-1}<\lambda_\ell\le c_t}
\mathbf u_\ell\mathbf u_\ell^\top$.
The filtration prioritizes smoother,
low-frequency treatment features
\citep{shuman2013}.
Repeated eigenvalues are grouped within
the same layer to ensure basis invariance.

\paragraph*{Haar and wavelet filtrations.}
Wavelet representations organize functions
from coarse to fine resolutions
\citep{mallat1989}.
Let $\{\boldsymbol\psi_{t,k}:
k\in\mathcal I_t\}$ be orthonormal
detail vectors in $\mathbb R^n$,
mutually orthogonal across resolutions
and orthogonal to $\mathbf1_n$.
Define
$\mathcal S_t^{\mathrm{wav}}
=\operatorname{span}\{\boldsymbol\psi_{t,k}:
k\in\mathcal I_t\}$,
$\mathcal H_t^{\mathrm{wav}}
=\operatorname{span}\{\mathbf1_n\}
\oplus\bigoplus_{r=1}^t
\mathcal S_r^{\mathrm{wav}}$,
and
$\mathbf B_t^{\mathrm{wav}}
=\sum_{k\in\mathcal I_t}
\boldsymbol\psi_{t,k}
\boldsymbol\psi_{t,k}^\top$.
For Haar filtrations, detail vectors
represent contrasts at successive
resolutions.
On base-$s$ rectangular grids,
tensor-product Haar details can be
organized by refinement depth and
projection dimension, reflecting the
stratification priorities underlying
the Tian--Xu and Shi--Xu criteria.

\begin{figure}[tbp]
\centering
\includegraphics[
    width=\textwidth,
    keepaspectratio
]{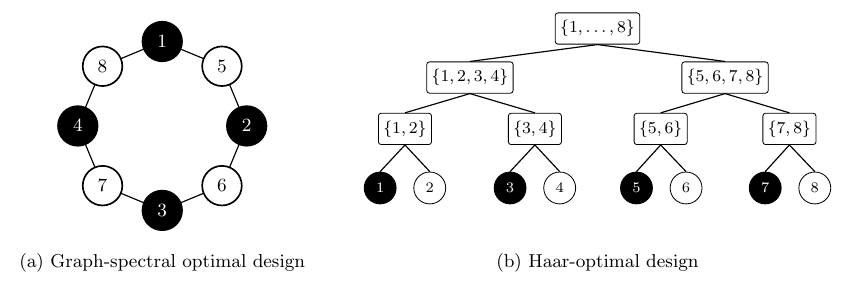}
\caption{Optimal four-point designs under
graph-spectral and Haar filtrations
on eight candidate treatments.
(a) A graph-spectral optimal design
$S_G=\{1,2,3,4\}$ for the eight-cycle
$1-5-2-6-3-7-4-8-1$.
(b) A Haar-optimal design
$S_H=\{1,3,5,7\}$ for the
three-level binary partition.
Black nodes indicate selected treatments.
The two filtrations have no common
lexicographically optimal four-point design.}
\label{fig:graph-wavelet-example}
\end{figure}

\paragraph*{Illustration.}
Consider eight candidate treatments and
four experimental runs without replication,
with $\mathbf V=\mathbf J_4$.
The graph-spectral filtration uses an
unweighted eight-cycle with vertex order
$1-5-2-6-3-7-4-8-1$.
Its three nonconstant layers correspond
to Laplacian eigenvalues
$\{2-\sqrt{2}\}$, $\{2\}$, and
$\{2+\sqrt{2},4\}$.
The Haar filtration uses successive binary
partitions into consecutive halves, pairs,
and singletons, giving three layers.
Among the $\binom{8}{4}=70$ possible
four-point designs, lexicographic
minimization under the graph-spectral
filtration selects either
$\{1,2,3,4\}$ or $\{5,6,7,8\}$.
Under the Haar filtration, the optimal
designs select one treatment from each
of the four pairs
$\{1,2\}$, $\{3,4\}$,
$\{5,6\}$, and $\{7,8\}$.
For the representative optimal designs
$S_G=\{1,2,3,4\}$ and
$S_H=\{1,3,5,7\}$,
the corresponding score vectors are
$\mathbf K^G(S_G)=(0,0,1/8)$ and
$\mathbf K^{\mathrm{wav}}(S_H)=(0,0,1/8)$.
In contrast,
$\mathbf K^G(S_H)=(0,1/8,0)$ and
$\mathbf K^{\mathrm{wav}}(S_G)=(1/8,0,0)$.
Figure~\ref{fig:graph-wavelet-example}
illustrates the two optimal designs.
Thus, the two filtrations have disjoint
sets of lexicographically optimal designs,
demonstrating that treatment-feature
priorities can fundamentally change
optimal treatment selection.

\FloatBarrier
\section{MMD Interpretation and RKHS Bounds}
\label{app:mmd-results}
\label{app:mmd-identities}

We show that F-OMA scores admit exact
maximum mean discrepancy (MMD) representations,
providing a kernel-discrepancy interpretation
of treatment-feature alignment with
experimental-unit structures.

Let
$\mu_{\mathcal X}
=n^{-1}\sum_{a=1}^n
\delta_{\mathbf x^{(a)}}$
denote the uniform distribution on
the candidate treatment space, and define
the empirical joint distribution
$P_d=N^{-1}\sum_{i=1}^N
\delta_{(\mathbf x_i,u_i)}$.
For any probability distribution $q_\Omega$
on the labeled experimental units, let
$Q_0=\mu_{\mathcal X}\otimes q_\Omega$.
For each filtration layer $t$, define
the product kernel
$k_{t,\mathbf V}
\{(\mathbf x,u_i),(\mathbf x',u_j)\}
=p_t(\mathbf x,\mathbf x')V_{ij}$.
For positive weights
$\mathbf a=(a_1,\ldots,a_T)^\top$, let
$k_{\mathbf a,\mathbf V}
=\sum_{t=1}^T a_t k_{t,\mathbf V}$,
and write $k=k_{\mathbf a,\mathbf V}$.
These kernels are positive semidefinite
because the treatment and unit kernels
are positive semidefinite.

\begin{proposition}[MMD representations]
\label{prop:app-foma-mmd}
For any design $d$ and
$\mathbf V\succeq0$, each F-OMA
layer score satisfies
$K_t(d;\mathbf V)
=\operatorname{MMD}_{k_{t,\mathbf V}}
(P_d,Q_0)^2$.
Consequently,
$A_{\mathbf a}(d;\mathbf V)
=\operatorname{MMD}_{k}(P_d,Q_0)^2$.
In particular, for
$\mathbf V=\mathbf J_N$,
$K_t(d;\mathbf J_N)
=\operatorname{MMD}_{p_t}
(\nu_d,\mu_{\mathcal X})^2$,
where
$\nu_d=N^{-1}\sum_{i=1}^N
\delta_{\mathbf x_i}$.
Moreover, every $g$ in the RKHS
$\mathcal H_k$ of $k$ satisfies
$|N^{-1}\sum_{i=1}^N
g(\mathbf x_i,u_i)-\int g\,dQ_0|
\leq
\|g\|_{\mathcal H_k}
\sqrt{A_{\mathbf a}(d;\mathbf V)}$.
\end{proposition}

\begin{proof}
Since each filtration increment space is
orthogonal to constants,
$\mathbf B_t\mathbf1_n=\mathbf0$.
Hence
$\int p_t(\mathbf x,\mathbf x')
\,d\mu_{\mathcal X}(\mathbf x')=0$
for every $\mathbf x\in\mathcal X$.
It follows that
$\int k_{t,\mathbf V}(\mathbf z,\mathbf z')
\,dQ_0(\mathbf z')=0$
for every $\mathbf z\in\mathcal X\times\Omega$.
Thus, the cross and reference terms
in the squared MMD vanish, giving
$\operatorname{MMD}_{k_{t,\mathbf V}}
(P_d,Q_0)^2
=N^{-2}\sum_{i,j=1}^N
p_t(\mathbf x_i,\mathbf x_j)V_{ij}
=K_t(d;\mathbf V)$.
Linearity in the kernel yields the
weighted representation.
When $\mathbf V=\mathbf J_N$,
the kernel depends only on treatments,
giving the stated treatment-distribution
representation.
Finally, the integration bound follows
from the definition of MMD
\citep{gretton2012} and the
Cauchy--Schwarz inequality in $\mathcal H_k$.
\end{proof}

The layerwise MMD representation also
clarifies the variance interpretation
in Theorem~\ref{thm:foma-variance}.
Under the specified random-coefficient
model, each interaction-layer variance is
$\operatorname{Var}(\overline\zeta_{t,d})
=\tau_t^2
\operatorname{MMD}_{k_{t,\mathbf V}}
(P_d,Q_0)^2$.
Choosing $a_t=\tau_t^2$ therefore
makes the weighted MMD criterion equal
to the design-dependent variance term
in \eqref{eq:response-level-variance}.
Thus, kernel discrepancy and variance
reduction provide equivalent interpretations
under this model.

The reference distribution $q_\Omega$
is arbitrary because the product kernels
have zero mean under $\mu_{\mathcal X}$.
Consequently, this representation
measures discrepancies involving centered
treatment features and unit relationships,
rather than general discrepancies between
joint probability distributions.
In particular, functions in $\mathcal H_k$
have zero treatment average at each unit,
so nonzero treatment-independent unit effects
are not covered by the RKHS bound.

\section{Layer-Weighted GLS Prediction under
an Estimable Filtration}
\label{app:global-prediction}

We relate F-OMA scores to prediction error
under a specified covariance model, allowing
different treatment layers to have different
prediction priorities.
Define $\mathbf q_0=n^{-1/2}\mathbf1_n$,
$\mathbf Q_R=[\mathbf q_0,\mathbf Q_1,\ldots,
\mathbf Q_{T_0}]$, and
$\mathbf Z_R=\mathbf X_d\mathbf Q_R$,
with $r=1+\sum_{t=1}^{T_0}r_t$.
Assume $\operatorname{rank}(\mathbf Z_R)=r\le N$,
ensuring estimability of the retained
treatment-function coefficients.
Consider the model
$\mathbf f=\mathbf Q_R\boldsymbol\beta$
and
$\mathbf Y=\mathbf Z_R\boldsymbol\beta
+\boldsymbol\varepsilon$,
where $\mathbf f$ contains the mean responses
over $\mathcal X$,
$\mathbb E(\boldsymbol\varepsilon)=\mathbf0$,
and $\operatorname{Cov}(\boldsymbol\varepsilon)
=\sigma^2\mathbf V$,
with $\sigma^2>0$ and $\mathbf V\succ0$.
Here $\mathbf V$ is the residual covariance
matrix up to scale, rather than merely
a design kernel, and is treated as known.
Let $\mathbf M=\mathbf Z_R^\top
\mathbf V^{-1}\mathbf Z_R$.
The generalized least squares (GLS)
estimator is
$\widehat{\boldsymbol\beta}
=\mathbf M^{-1}\mathbf Z_R^\top
\mathbf V^{-1}\mathbf Y$,
giving $\widehat{\mathbf f}
=\mathbf Q_R\widehat{\boldsymbol\beta}$.

\paragraph*{Layer-weighted prediction error.}

For positive weights
$\mathbf a=(a_0,\ldots,a_{T_0})^\top$,
define
$\mathbf D_{\mathbf a}
=\operatorname{diag}
(a_0,a_1\mathbf I_{r_1},\ldots,
a_{T_0}\mathbf I_{r_{T_0}})$.
The weights specify prediction priorities
for the retained treatment layers.
Define the layer-weighted integrated
mean squared prediction error by
\[
\begin{aligned}
\operatorname{IMSE}_{\mathbf a}(d)
=\frac1n\mathbb E\bigl\{
&\|\mathbf D_{\mathbf a}^{1/2}
\mathbf Q_R^\top
(\widehat{\mathbf f}-\mathbf f)\|_2^2\\
&+\|(\mathbf I_n-\mathbf Q_R\mathbf Q_R^\top)
(\widehat{\mathbf f}-\mathbf f)\|_2^2
\bigr\}.
\end{aligned}
\]
The first term weights errors in the retained
layers, while the second measures errors
orthogonal to the retained space.
Let $\mathbf G=\mathbf Z_R^\top\mathbf Z_R$
and define the weighted coverage factor
\[
\gamma_{\mathbf a}(d)
=\left\{
N\left\|
\mathbf D_{\mathbf a}^{1/2}
\mathbf G^{-1}
\mathbf D_{\mathbf a}^{-1/2}
\right\|_2
\right\}^{-1}>0.
\]
The constant treatment feature contributes
$K_0(\mathbf V)
=\mathbf1_N^\top\mathbf V\mathbf1_N/(nN^2)$.

\begin{theorem}[Layer-weighted GLS prediction bound]
\label{thm:global-prediction}
Under the stated model and estimability
condition,
$\operatorname{IMSE}_{\mathbf a}(d)
=(\sigma^2/n)
\operatorname{tr}
(\mathbf D_{\mathbf a}\mathbf M^{-1})$,
and
\begin{equation}
\label{eq:global-prediction-bound}
\operatorname{IMSE}_{\mathbf a}(d)
\le
\frac{\sigma^2}
{n\gamma_{\mathbf a}(d)^2}
\left\{
a_0K_0(\mathbf V)
+\sum_{t=1}^{T_0}
a_tK_t(d;\mathbf V)
\right\}.
\end{equation}
\end{theorem}

\begin{proof}
Since $\mathbf f$ and
$\widehat{\mathbf f}$ belong to
$\operatorname{col}(\mathbf Q_R)$,
the prediction error has no component
orthogonal to the retained space.
Using the covariance of the GLS estimator,
we obtain
$\operatorname{IMSE}_{\mathbf a}(d)
=(\sigma^2/n)
\operatorname{tr}
(\mathbf D_{\mathbf a}\mathbf M^{-1})$.
Let $\mathbf C=\mathbf Z_R^\top
\mathbf V\mathbf Z_R$.
Both $\mathbf G$ and $\mathbf C$ are
positive definite.
The matrix inequality
$\mathbf M\succeq
\mathbf G\mathbf C^{-1}\mathbf G$
follows by expanding
$\mathbf R^\top\mathbf R\succeq\mathbf0$,
where
$\mathbf R=\mathbf V^{-1/2}\mathbf Z_R
-\mathbf V^{1/2}\mathbf Z_R
\mathbf C^{-1}\mathbf G$.
Consequently,
$\mathbf M^{-1}\preceq
\mathbf G^{-1}\mathbf C\mathbf G^{-1}$.
Set
$\mathbf B=\mathbf D_{\mathbf a}^{1/2}
\mathbf G^{-1}\mathbf D_{\mathbf a}^{-1/2}$
and
$\mathbf C_{\mathbf a}
=\mathbf D_{\mathbf a}^{1/2}
\mathbf C\mathbf D_{\mathbf a}^{1/2}$.
Since $\mathbf C_{\mathbf a}\succeq\mathbf0$,
\[
\begin{aligned}
\operatorname{tr}
(\mathbf D_{\mathbf a}\mathbf M^{-1})
&\le
\operatorname{tr}
(\mathbf B\mathbf C_{\mathbf a}\mathbf B^\top)\\
&\le
\|\mathbf B\|_2^2
\operatorname{tr}(\mathbf C_{\mathbf a})\\
&=
\frac{\operatorname{tr}
(\mathbf D_{\mathbf a}\mathbf C)}
{N^2\gamma_{\mathbf a}(d)^2}.
\end{aligned}
\]
Finally, orthogonality of the retained
treatment layers gives
$\operatorname{tr}
(\mathbf D_{\mathbf a}\mathbf C)/N^2
=a_0K_0(\mathbf V)
+\sum_{t=1}^{T_0}a_tK_t(d;\mathbf V)$,
proving the bound.
\end{proof}

\begin{corollary}[Fixed treatment multiset]
\label{cor:fixed-multiset-prediction}
For a fixed unit kernel and treatment
multiset satisfying estimability,
$\gamma_{\mathbf a}(d)$ and
$K_0(\mathbf V)$ are invariant under
unit reassignment.
Thus, minimizing
$\sum_{t=1}^{T_0}a_tK_t(d;\mathbf V)$
over these assignments also minimizes
the upper bound in
\eqref{eq:global-prediction-bound}.
In particular, let
$a_t=\epsilon^{t-1}$ for
$t=1,\ldots,T_0$, with sufficiently
small $\epsilon>0$.
Then any globally lexicographically
optimal assignment for the retained
F-OMA layers minimizes this bound.
\end{corollary}

\begin{proof}
Reassigning a fixed treatment multiset
only permutes the rows of $\mathbf Z_R$,
leaving $\mathbf G$ and hence
$\gamma_{\mathbf a}(d)$ unchanged.
Since $K_0(\mathbf V)$ is also
assignment-independent, the first
conclusion follows from
\eqref{eq:global-prediction-bound}.
The feasible assignment class is finite.
For sufficiently small $\epsilon>0$,
minimizing
$\sum_{t=1}^{T_0}\epsilon^{t-1}
K_t(d;\mathbf V)$
is equivalent to lexicographically
minimizing the retained score vector.
The second conclusion follows.
\end{proof}

The corollary provides a prediction-bound
interpretation of F-OMA priorities under
fixed treatment multisets.
It does not establish minimization of
actual prediction error, and the bound
may be loose for strongly unequal weights.

\paragraph*{Approximation and aliasing.}

If the retained filtration does not fully
represent the mean response, write
$\mathbf f=\mathbf Q_R\boldsymbol\beta
+\mathbf h$, where
$\mathbf Q_R^\top\mathbf h=\mathbf0$.
The observed responses then satisfy
$\mathbf Y=\mathbf Z_R\boldsymbol\beta
+\mathbf X_d\mathbf h
+\boldsymbol\varepsilon$.
Define
$\mathbf b_{\mathbf h}
=\mathbf M^{-1}\mathbf Z_R^\top
\mathbf V^{-1}\mathbf X_d\mathbf h$.
The prediction error decomposes as
\begin{equation*}
\operatorname{IMSE}_{\mathbf a}(d)
=\frac1n\left\{
\|\mathbf h\|_2^2
+\mathbf b_{\mathbf h}^\top
\mathbf D_{\mathbf a}\mathbf b_{\mathbf h}
+\sigma^2\operatorname{tr}
(\mathbf D_{\mathbf a}\mathbf M^{-1})
\right\}.
\end{equation*}
The three terms represent approximation
error, weighted aliasing error, and
weighted estimation variance, respectively.
The bound in
\eqref{eq:global-prediction-bound}
controls only the final term.

\section{Constrained Construction of F-OMA Designs}
\label{app:construction}

We consider the construction of F-OMA designs
without treatment replication and with exact
one-factor marginal balance.
Let $\mathcal X_c\subseteq\mathcal X
=\mathcal X_1\times\cdots\times\mathcal X_m$
be a finite candidate pool of size $M\ge N$,
where $|\mathcal X_j|=s_j$ and $m\ge2$.
Assume that each $s_j$ divides $N$ and that
the following feasible design class is nonempty:
\begin{equation*}
\mathcal D_{\mathrm{bal}}
=\left\{
d=(\mathbf x_1,\ldots,\mathbf x_N)
\in\mathcal X_c^N:
\begin{array}{l}
\mathbf x_i\ne\mathbf x_k
\quad(i\ne k),\\
\displaystyle
\sum_{i=1}^N\mathbb I\{x_{ij}=\ell\}
=N/s_j\\
\quad(j=1,\ldots,m;\
\ell\in\mathcal X_j)
\end{array}
\right\},
\end{equation*}
where $\mathbf x_i=(x_{i1},\ldots,x_{im})^\top$
is assigned to the labeled unit $u_i$.
Because a one-point replacement generally
violates marginal balance, we use two
constraint-preserving moves.

\paragraph*{Margin-preserving configuration exchange.}
Choose $i<j$ and a nonempty proper subset
$J\subset[m]=\{1,\ldots,m\}$.
Replace $(\mathbf x_i,\mathbf x_j)$ by
$(\mathbf y,\mathbf z)$, where
$(y_k,z_k)=(x_{jk},x_{ik})$ for $k\in J$
and $(y_k,z_k)=(x_{ik},x_{jk})$ otherwise.
The move is admissible if
$\mathbf y,\mathbf z\in\mathcal X_c$,
$\mathbf y\ne\mathbf z$, and neither
offspring duplicates another selected
configuration.
Since the unordered pair of levels for
each factor is unchanged, all one-factor
margins are preserved.
Moves that leave the design unchanged or
coincide with an assignment swap are omitted.

\paragraph*{Unit-assignment swap.}
Interchanging $\mathbf x_i$ and $\mathbf x_j$
between their labeled units preserves
the selected treatment multiset and hence
both constraints.
Thus, configuration exchanges modify
treatment selection, whereas assignment
swaps modify treatment--unit assignment.

\paragraph*{Feasible starting designs.}
A feasible configuration set can be obtained
by solving a binary feasibility problem
with $z_{\mathbf x}\in\{0,1\}$ for
$\mathbf x\in\mathcal X_c$, subject to
$\sum_{\mathbf x\in\mathcal X_c}z_{\mathbf x}=N$
and
$\sum_{\mathbf x\in\mathcal X_c:\,x_j=\ell}
z_{\mathbf x}=N/s_j$
for every $j$ and $\ell\in\mathcal X_j$.
Random linear objectives can be used to
generate diverse feasible sets, which are
then randomly assigned to the labeled units.
The cost of generating feasible starts is
excluded from the local-search analysis below.

Let $\widetilde{\mathcal J}$ denote the family
of coordinate subsets considered for
configuration exchanges.
Algorithm~\ref{alg:foma-exchange} performs
multistart lexicographic optimization using
these two neighborhoods.
Ties in move selection and comparisons
across starts are resolved using fixed
deterministic rules.

\begin{algorithm}[t]
\small
\DontPrintSemicolon
\caption{Balanced no-replication multistart F-OMA construction}
\label{alg:foma-exchange}

\KwIn{$\mathcal X_c$, $\mathbf V$,
$\{p_t\}_{t=1}^T$,
$\widetilde{\mathcal J}
\subseteq2^{[m]}\setminus
\{\varnothing,[m]\}$,
number of starts $S$}

\KwOut{$\widehat d\in\mathcal D_{\mathrm{bal}}$}

\For{$s=1,\ldots,S$}{
  Generate $d^{(s)}\in\mathcal D_{\mathrm{bal}}$
  and store its selected set\;

  \Repeat{no improving move exists}{
    Evaluate all admissible configuration
    exchanges with
    $J\in\widetilde{\mathcal J}$
    and all assignment swaps\;

    \If{a lexicographically improving move exists}{
      Apply the move with the smallest
      lexicographic score change and
      update the selected set\;
    }
  }
}

\textbf{return} the lexicographically best
terminal design $\widehat d$\;
\end{algorithm}

\FloatBarrier

\paragraph*{Exact score updates and convergence.}
For a real symmetric treatment kernel $h$,
define
$A_h(d;\mathbf V)
=N^{-2}\sum_{r,s=1}^N
V_{rs}h(\mathbf x_r,\mathbf x_s)$.
In particular, $A_{p_t}(d;\mathbf V)
=K_t(d;\mathbf V)$.
For brevity, write
$h_{ik}=h(\mathbf x_i,\mathbf x_k)$,
$h_{\mathbf y,k}=h(\mathbf y,\mathbf x_k)$,
and similarly for other subscripts.

\begin{proposition}[Exact updates and finite descent]
\label{prop:exchange-construction}
\label{prop:constrained-exchange-updates}
For any real symmetric $h$ and $\mathbf V$,
an admissible configuration exchange
$(\mathbf x_i,\mathbf x_j)
\mapsto(\mathbf y,\mathbf z)$
has exact increment
\begin{equation}
\label{eq:balanced-exchange-update}
\begin{aligned}
\Delta_{ij,J}^{\mathrm{bal}}(h)
=\frac{1}{N^2}\Biggl[
&2\sum_{k\notin\{i,j\}}
\Bigl\{
V_{ik}(h_{\mathbf y,k}-h_{ik})\\
&\qquad+
V_{jk}(h_{\mathbf z,k}-h_{jk})
\Bigr\}\\
&+2V_{ij}
(h_{\mathbf y,\mathbf z}-h_{ij})\\
&+V_{ii}
(h_{\mathbf y,\mathbf y}-h_{ii})\\
&+V_{jj}
(h_{\mathbf z,\mathbf z}-h_{jj})
\Biggr].
\end{aligned}
\end{equation}
For an assignment swap, the increment is
\begin{equation}
\label{eq:assignment-swap-update}
\begin{aligned}
\Delta_{ij}^{\mathrm{sw}}(h)
=\frac{1}{N^2}\Biggl[
&2\sum_{k\notin\{i,j\}}
(V_{ik}-V_{jk})
(h_{jk}-h_{ik})\\
&+(V_{ii}-V_{jj})
(h_{jj}-h_{ii})
\Biggr].
\end{aligned}
\end{equation}
Each increment requires $O(N)$ operations
under constant-time kernel lookup, so all
$T$ layerwise increments require $O(TN)$,
excluding feasibility checks.
Every admissible move preserves
$\mathcal D_{\mathrm{bal}}$.
On this finite design class, strict
lexicographic descent terminates at a
local optimum relative to the searched
exchange and assignment-swap neighborhoods,
provided that the final neighborhood
scan is exhaustive.
\end{proposition}

\begin{proof}
Under a configuration exchange, only terms
involving runs $i$ or $j$ change in
$N^2A_h(d;\mathbf V)$.
Symmetry combines the off-diagonal terms
involving another run $k$ into the first
sum of \eqref{eq:balanced-exchange-update}.
The remaining cross and diagonal terms
give the other contributions.
Setting $\mathbf y=\mathbf x_j$ and
$\mathbf z=\mathbf x_i$ yields
\eqref{eq:assignment-swap-update}.
Both formulas contain $O(N)$ terms and
apply separately to every $p_t$.

Configuration exchanges preserve marginal
counts, while their admissibility conditions
prevent replication.
Assignment swaps preserve the selected
treatment multiset.
Thus, all accepted moves remain feasible.
Each accepted move strictly decreases
$\mathbf K(d;\mathbf V)$ lexicographically.
Since $\mathcal D_{\mathrm{bal}}$ is finite,
the algorithm terminates after finitely
many moves.
A complete terminal scan with no improving
move establishes local optimality relative
to the searched neighborhoods.
\end{proof}

The diagonal terms in these updates cannot
generally be omitted, since treatment kernels
need not have constant diagonals.
When $\mathbf V=\mathbf J_N$, every
assignment-swap increment vanishes because
the criterion depends only on the selected
treatment multiset.

\paragraph*{Scalarized search.}
The same updates apply to
$A_{\mathbf a}(d;\mathbf V)
=\sum_{t=1}^T a_tK_t(d;\mathbf V)$
by taking
$h=h_{\mathbf a}
=\sum_{t=1}^T a_tp_t$.
For a numerical tolerance $\tau\ge0$,
a scalarized search may accept moves whose
increments are below $-\tau$.
A complete terminal scan then guarantees
$A_{\mathbf a}(d^\star;\mathbf V)
\le A_{\mathbf a}(d';\mathbf V)+\tau$
for every searched admissible neighbor $d'$.
Exact local optimality follows when $\tau=0$.
By Proposition~\ref{prop:foma-scalarization},
sufficiently small weights
$a_t=\epsilon^{t-1}$ preserve the
lexicographic ordering on the finite
design class.
Direct lexicographic comparisons avoid
choosing such an $\epsilon$ numerically.
Exact local-optimality guarantees require
strict comparisons and a complete terminal
scan; numerical tolerances or early stopping
do not provide the same guarantee.

\paragraph*{Computational complexity.}

\begin{proposition}[Computational complexity]
\label{prop:construction-complexity}
Let $L=|\widetilde{\mathcal J}|$.
Suppose candidate membership and selected-set
membership use hash tables with expected
constant-time lookup after configuration
encoding, and constructing both offspring
costs $O(m)$.
Assume also that the required treatment-kernel
values are available in constant time.
Then one exhaustive scalarized neighborhood
scan has expected cost
$O\{LN^2(N+m)+N^3\}$.
Direct lexicographic search has expected
scan cost
$O\{LN^2(m+TN)+TN^3\}$.
For $S$ starts and at most $I$ accepted
moves per start, the respective total
local-search costs are obtained by
multiplying these bounds by $S(I+1)$.
\end{proposition}

\begin{proof}
Each scan considers at most
$\binom N2 L$ configuration exchanges
and $\binom N2$ assignment swaps.
Constructing and checking the offspring
costs expected $O(m)$ per exchange,
while the scalar score update costs $O(N)$.
Each assignment swap costs $O(N)$.
This gives the scalarized scan bound.
For direct lexicographic search, evaluating
all $T$ layer increments costs $O(TN)$
per move, giving the second bound.
At most $I$ accepted moves and one terminal
scan are performed per start.
\end{proof}

For the full subset family,
$L=2^m-2$, so the scan cost is $O(N^3)$
when $m$ and $T$ are fixed.
Although finite termination is guaranteed,
no polynomial bound on the number of
accepted moves or global optimality is claimed.
For large $m$, one may restrict the exchange
family to
$\widetilde{\mathcal J}_q
=\{J\subset[m]:1\le|J|\le q\}$,
where $1\le q\le m-1$ and
$|\widetilde{\mathcal J}_q|
=\sum_{r=1}^q\binom mr$.
This preserves feasibility, but local
optimality then refers only to the
restricted neighborhood.

The complexity bounds exclude feasible-start
generation and kernel preprocessing.
If the layer kernels are already available,
precomputing $h_{\mathbf a}$ costs
$O(TM^2)$ operations.
The scalarized kernel table and unit kernel
require $O(M^2+N^2)$ storage, while retaining
all $T$ layer kernels requires $O(TM^2)$
treatment-kernel storage.
Candidate coordinates and the subset family
require an additional $O(Mm+Lm)$ storage.

\section{Simulation and Bayesian Optimization Details}
\label{app:simulation-details}

This appendix specifies the response generator and implementation for
Section~\ref{sec:numerical-studies}.

\paragraph*{Treatment surface and response generation.}

Write $n=72$ and $N=36$. The scaled coordinates
$\mathbf z=(a,b,w,o,\delta)^\top$ represent learning rate, batch size,
hidden width, optimizer, and dropout; the first two take values
$0,1/2,1$ and the others $0,1$. Define
\[
\begin{aligned}
f_{\mathrm{raw}}(\mathbf{z})
={}&1.30(a-0.68)^2
+0.95(b-0.35)^2
+0.30(w-0.70)^2\\
&+0.58\sin(2.7\pi a+0.6b)
+0.42\cos(1.7\pi b+0.8\delta)\\
&+0.55(a-b)^2(1+0.7w)
+0.70o(a-0.25)(b+0.30)\\
&+0.35w\delta-0.32ow
+0.33(1-o)\cos(3.2\pi ab)\\
&+0.27\sin\{2.3\pi(a+0.35b+0.12w-0.18\delta)\}\\
&-1.45\exp[-10\{(a-0.58)^2+(b-0.28)^2\}]
(1+0.18w-0.25\delta).
\end{aligned}
\]
Let $f$ be its centered, unit-RMS version over all 72 candidates.
In \eqref{eq:simulation-response}, set
$R(\mathbf x)=10+f(\mathbf x)$,
$g_b(u)=2\sqrt{1-\rho}\,b_b(u)$,
$h_b(\mathbf x,u)=2\sqrt{\rho}\,H_b(\mathbf x,u)$,
and $\rho=0.8$, with independent $N(0,0.15^2)$ observation errors
within each trajectory.

The six raw interaction features are
\[
\begin{aligned}
\psi_1(\mathbf z)&=a,
&
\psi_2(\mathbf z)&=\sin(1.4\pi b+0.2),\\
\psi_3(\mathbf z)&=(a-0.4)(b-0.6),
&
\psi_4(\mathbf z)&=(2o-1)(a-0.25),\\
\psi_5(\mathbf z)&=w\delta+0.35w-0.2\delta,
&
\psi_6(\mathbf z)&=\cos(2.1\pi a+1.2b)+0.3\delta o.
\end{aligned}
\]
For the scaled coordinates $\mathbf z^{(j)}$ of candidate $\mathbf x^{(j)}$,
define, for
$\ell=1,\ldots,6$,
\[
\begin{gathered}
\bar\psi_\ell=\frac1n\sum_{j=1}^{n}\psi_\ell(\mathbf z^{(j)}),
\qquad
s_\ell^2=\frac1n\sum_{j=1}^{n}
\{\psi_\ell(\mathbf z^{(j)})-\bar\psi_\ell\}^2,\\
\Phi_{j\ell}=
\frac{\psi_\ell(\mathbf z^{(j)})-\bar\psi_\ell}{s_\ell}.
\end{gathered}
\]
The columns of $\boldsymbol\Phi\in\mathbb R^{n\times6}$ have zero mean
and unit RMS over the complete candidate space but need not be orthogonal.

Independently across replicates and of the observation errors, draw
\[
\mathbf c_{b,0},\ldots,\mathbf c_{b,6}
\stackrel{\mathrm{ind}}{\sim}\mathcal N(\mathbf0,\mathbf K_u).
\]
With
\[
\mathbf A_b=(\mathbf c_{b,1},\ldots,\mathbf c_{b,6})
\in\mathbb R^{N\times6},
\qquad
\mathbf C_k=\mathbf I_k-\frac1k\mathbf1_k\mathbf1_k^\top,
\quad k\in\{n,N\},
\]
center and normalize each realization separately:
\[
\begin{aligned}
\mathbf b_b^\circ&=\mathbf C_N\mathbf c_{b,0},
&
\mathbf H_b^\circ&=\mathbf C_n\boldsymbol\Phi\mathbf A_b^\top\mathbf C_N,\\
\mathbf b_b&=\frac{\sqrt N\,\mathbf b_b^\circ}{\|\mathbf b_b^\circ\|_2},
&
\mathbf H_b&=\frac{\sqrt{nN}\,\mathbf H_b^\circ}{\|\mathbf H_b^\circ\|_F}.
\end{aligned}
\]
Here $b_b(u_i)=(\mathbf b_b)_i$ and
$H_b(\mathbf x^{(j)},u_i)=(\mathbf H_b)_{ji}$; the denominators are
positive almost surely under the kernels below. Consequently,
\[
\begin{gathered}
\frac1N\sum_{i=1}^N b_b(u_i)=0,\qquad
\frac1N\sum_{i=1}^N H_b(\mathbf x,u_i)=0
\quad(\mathbf x\in\mathcal X),\\
\frac1n\sum_{j=1}^n H_b(\mathbf x^{(j)},u)=0
\quad(u\in\Omega).
\end{gathered}
\]
Both $b_b$ and $H_b$ have unit RMS under the uniform unit and
treatment--unit product measures, respectively. Thus $R$ is the
environment-averaged expected loss, and the combined heterogeneity
$g_b(u)+h_b(\mathbf x,u)$ has RMS $2$ under the uniform product measure,
compared with RMS $1$ for $f$. The treatment surface and noise level are
fixed across structures and replicates; the unit effects and interactions vary.

\paragraph*{Unit kernels and initial designs.}

The simulation and unit-aware GP kernels differ from the design kernels
$\mathbf V_{\mathrm{des}}$ in Section~\ref{sec:numerical-studies}.
With equal unit weights, they are
\[
\mathbf K_u^{\mathrm{NS}}
=\frac43(3\mathbf E_{F_1}+2\mathbf E_{F_2}+\mathbf E_{F_3}),
\qquad
\mathbf K_u^{\mathrm{RC}}
=\frac{36}{11}(\mathbf E_R+\mathbf E_C).
\]
Both are positive semidefinite, satisfy
$\mathbf K_u\mathbf1_{36}=\mathbf0$, and have mean diagonal one.
They specify Gaussian coefficient dependence before normalization, not
necessarily the covariance of the final normalized effects.

All initial designs use 36 distinct configurations and no response information.
F-OMA uses direct lexicographic search with margin-preserving configuration
exchanges and assignment swaps: each three-level factor level occurs 12 times
and each two-level factor level 18 times. One design per unit structure is
fixed across replicates. SRS independently samples without replacement in
each replicate, without enforcing balance. 
MaxProLHD$\to$grid starts from a continuous midpoint Latin hypercube
constructed using the maximum-projection criterion. In column $j$,
the entry with rank $r_{ij}\in\{1,\ldots,36\}$ is mapped to
$\lfloor(r_{ij}-1)s_j/36\rfloor$, where
$(s_1,\ldots,s_5)=(3,3,2,2,2)$. Duplicate configurations are removed
by within-column swaps, preserving one-factor margins, and the
resulting rows are randomly assigned to units.
Both baselines are regenerated in each replicate
and do not optimize treatment--unit alignment.

\paragraph*{GP covariance and fitting.}

Both models have an unknown constant mean. For scaled coordinates, let
$d_x^2(\mathbf z,\mathbf z')=\sum_{j\in\{1,2,3,5\}}(z_j-z_j')^2$ and
$r_x=\sqrt5d_x/\ell$. The treatment correlation is
\[
k_x(\mathbf z,\mathbf z')
=(1+r_x+r_x^2/3)e^{-r_x}
\exp\{-\lambda_{\mathrm{cat}}\mathbf1(o\ne o')\},
\]
using a common Mat\'ern-$5/2$ length scale for the four ordered coordinates
and a categorical correlation for optimizer. For observation indices $i,j$,
the unit-aware covariance is
\[
\begin{aligned}
C_{ij}
={}&v_f k_x(\mathbf x_i,\mathbf x_j)+v_u k_u(u_i,u_j)\\
&+v_{xu}k_x(\mathbf x_i,\mathbf x_j)k_u(u_i,u_j)+v_e\delta_{ij},
\end{aligned}
\]
where $k_u$ comes from $\mathbf K_u$,
$\ell>0$, $\lambda_{\mathrm{cat}}\ge0$, and
$v_f,v_u,v_{xu},v_e\ge0$. All $k_x$ evaluations use scaled coordinates,
including those written in terms of $\mathbf x$. The unit-blind model
sets $v_u=v_{xu}=0$. Since $\delta_{ij}$ indexes observations, repeated
evaluations have independent observation noise.

Fitting uses standardized observed responses, ordinary kriging, and bounded
restricted maximum likelihood. Hyperparameters are fitted after initialization
and every six BO steps, using two deterministic starts and L-BFGS-B with at
most 80 iterations per start; observations update the posterior between refits.
The unit-aware GP is given the simulator's
coefficient-kernel matrix $\mathbf K_u$,
including its relative stratum weights,
but not the realized latent effects.
Its treatment-kernel and variance parameters are
estimated from observations.
The comparisons are conditional on the specified
unit-kernel structures and relative stratum weights.

Optimization uses logarithmic parameters with bounds
\[
\begin{gathered}
0.10\le\ell\le4,\qquad 0.02\le\lambda_{\mathrm{cat}}\le8,\\
10^{-5}\le v_f,v_u,v_{xu}\le50,\qquad 10^{-5}\le v_e\le10.
\end{gathered}
\]
In the order $(\ell,\lambda_{\mathrm{cat}},v_f,v_u,v_{xu},v_e)$,
the initial starts are
\[
(0.65,1,0.75,0.20,0.20,0.12)
\quad\text{and}\quad
(0.30,1,0.75,0.40,0.10,0.30).
\]
The unit-blind model omits $v_u,v_{xu}$ from these bounds and starts,
keeping them zero. At later refits, the first start uses the previous fit,
with variance parameters rescaled to the current standardized-response scale
and clipped to the bounds; the second start is unchanged. For
$n_{\mathrm{obs}}$ observations, the initial diagonal jitter is
\[
10^{-10}\max\left\{1,
\frac{\operatorname{tr}(\mathbf C)}{n_{\mathrm{obs}}}\right\},
\]
where $\mathbf C$ is the observation covariance on the standardized-response
scale. Failed Cholesky factorization triggers a factor-$10$ increase,
with at most five increases.

\paragraph*{Sequential optimization and pairing.}

As in Section~\ref{sec:numerical-studies}, each trajectory has 36 initial
evaluations and 36 follow-up steps. A shared, response-independent random
permutation of the 36 units determines the next unit before treatment
selection. Knowledge gradient \citep{frazier2009} considers all 72 candidates,
allows treatment repetition, and uses the fitted posterior of the
environment-averaged target; the recommendation minimizes that posterior mean.
KG is evaluated by exact one-dimensional Gaussian-envelope integration,
conditional on the fitted hyperparameters. Numerical ties in acquisition
and recommendation are resolved using a response-independent random
priority order shared by all six methods within each replicate.
The true $R$ is used only for performance assessment, never acquisition
or recommendation. Within each replicate, all six methods share latent
effects, the unit schedule, and common random numbers. The unit-aware and
unit-blind GPs for a given initializer receive identical initial data;
their later evaluations may differ.

Recommendation regret and normalized AURC \eqref{eq:bo-aurc} are evaluated
at $H=10,18,36$. Paired differences use the 100 replicates and unadjusted
95\% paired $t$ intervals, conditional on the fixed treatment surface
and the prespecified F-OMA design for each unit structure.

\section{Additional Bayesian Optimization Results}
\label{app:additional-bo-results}

These diagnostics supplement Section~\ref{sec:numerical-studies} using
the same 100 paired replicates.

\paragraph*{Shorter-horizon AURC comparisons.}

\begin{figure}[htbp]
\centering
\begin{minipage}[t]{0.49\textwidth}
\centering
\includegraphics[width=\linewidth]
{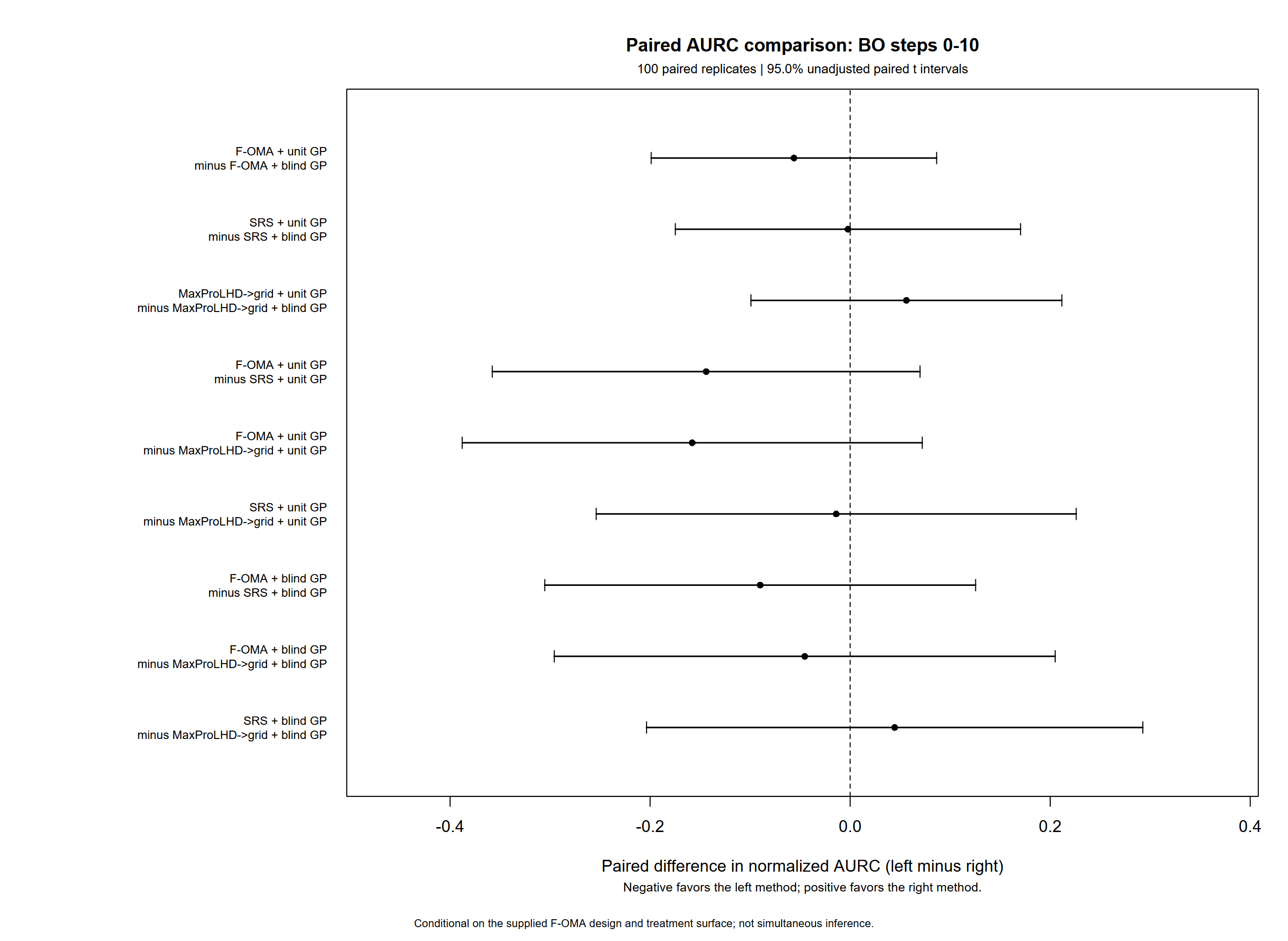}
\smallskip
{\small (a) Nested units}
\end{minipage}
\hfill
\begin{minipage}[t]{0.49\textwidth}
\centering
\includegraphics[width=\linewidth]
{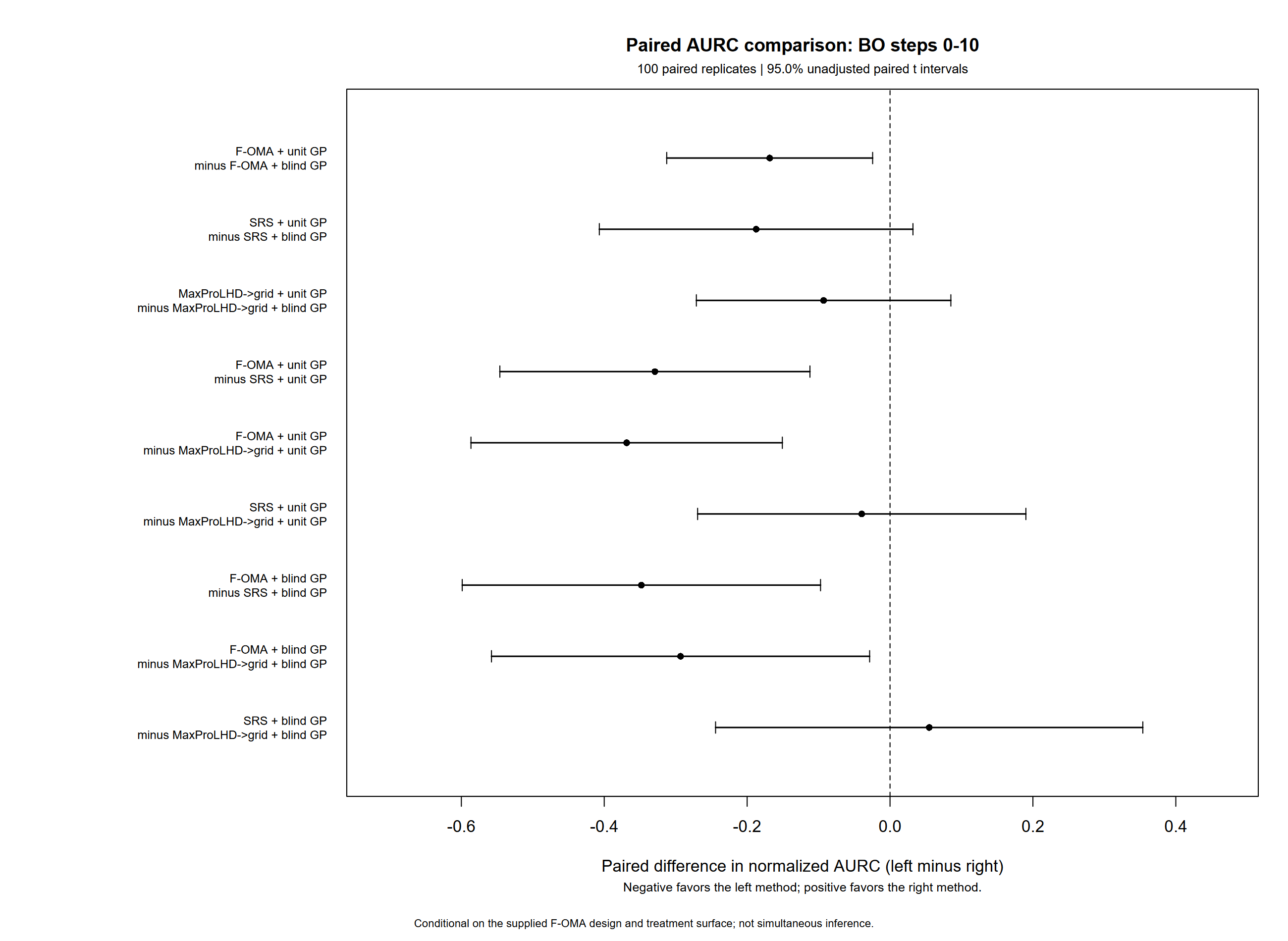}
\smallskip
{\small (b) Crossed units}
\end{minipage}
\caption{Paired differences in normalized AURC over BO steps 0--10.
Points show mean within-replicate differences, and bars show unadjusted
95\% paired $t$ intervals over 100 replicates.
Negative values favor the first method.}
\label{fig:bo-aurc-h10}
\end{figure}

\begin{figure}[htbp]
\centering
\begin{minipage}[t]{0.49\textwidth}
\centering
\includegraphics[width=\linewidth]
{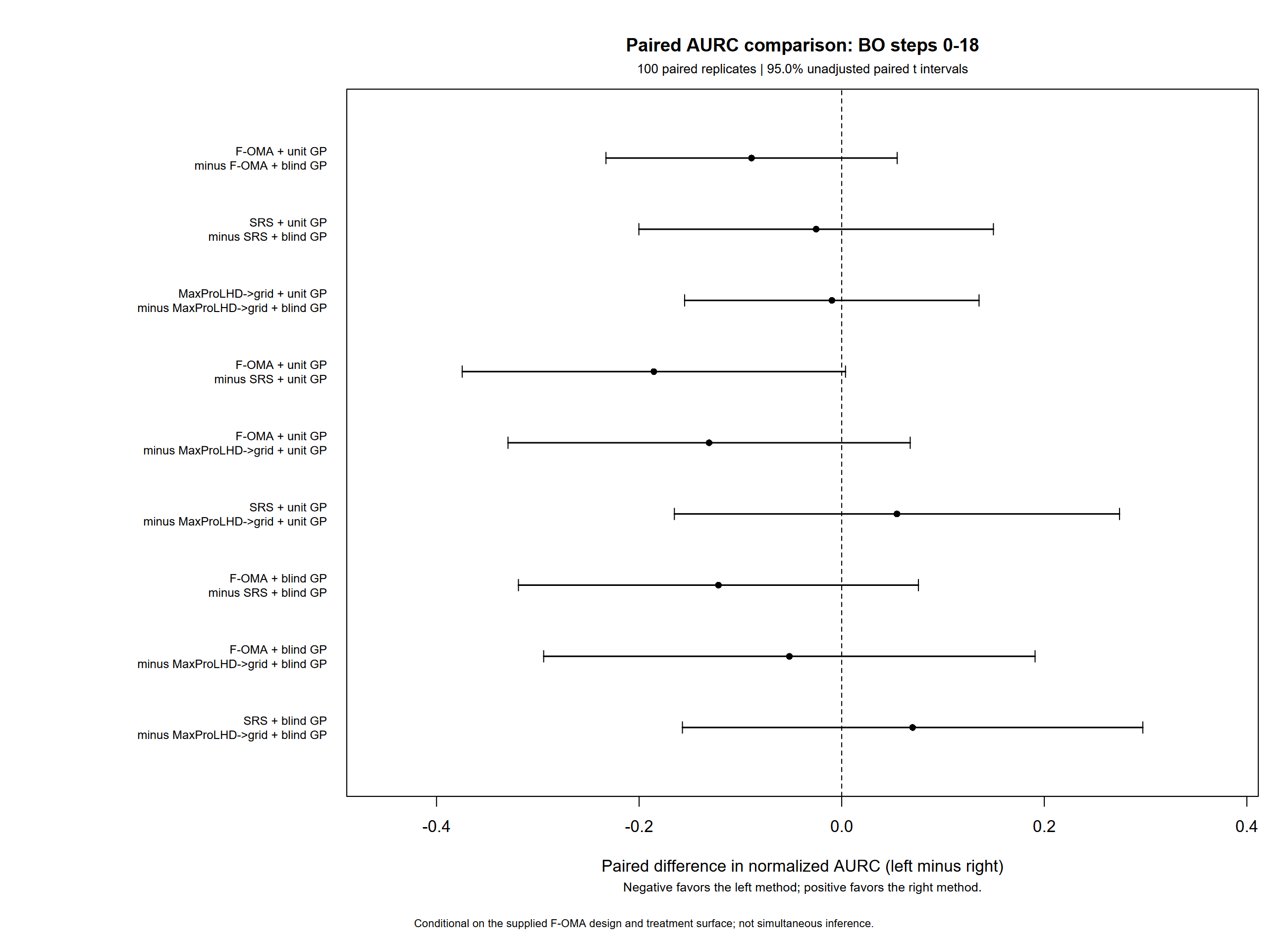}
\smallskip
{\small (a) Nested units}
\end{minipage}
\hfill
\begin{minipage}[t]{0.49\textwidth}
\centering
\includegraphics[width=\linewidth]
{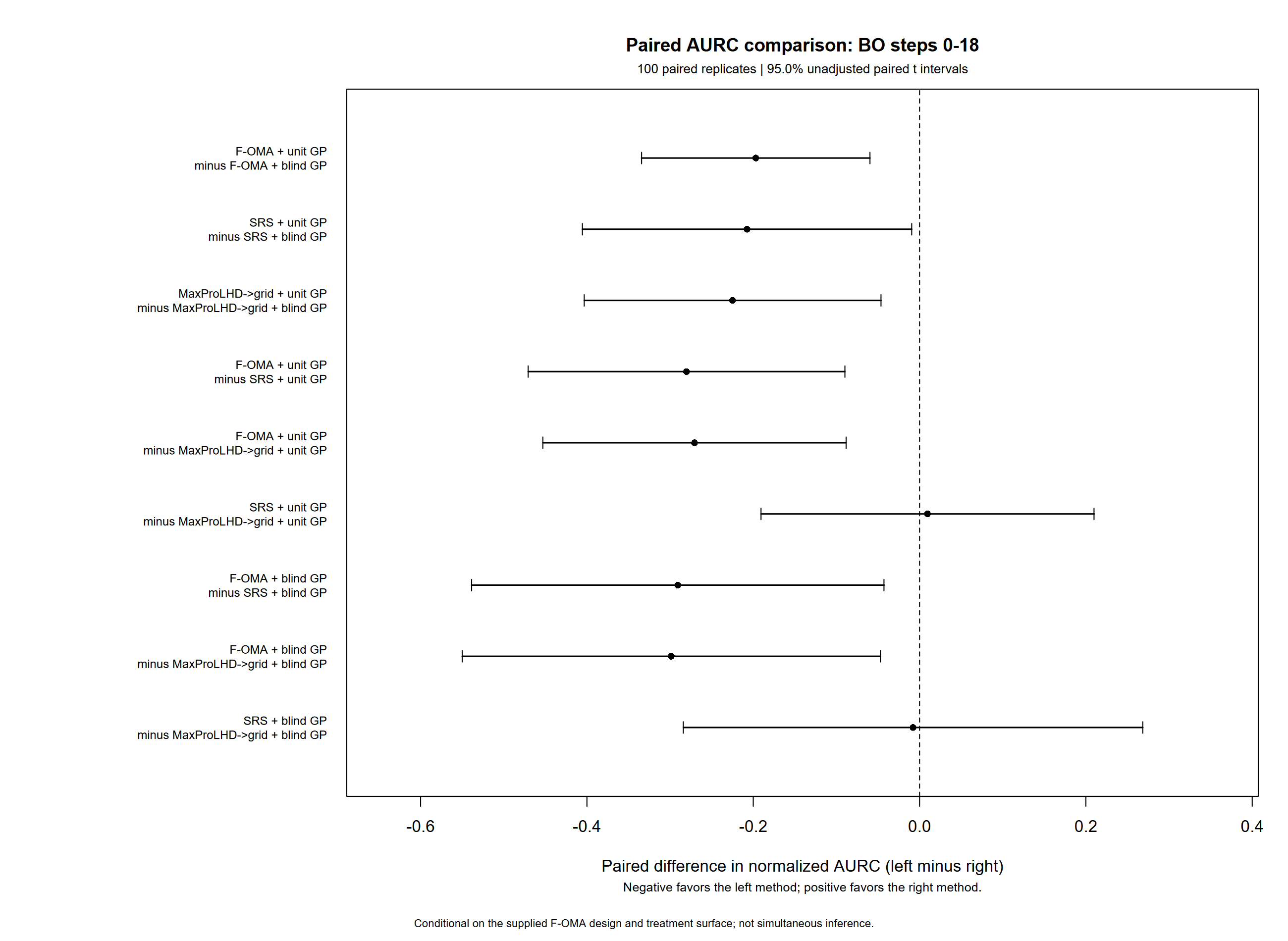}
\smallskip
{\small (b) Crossed units}
\end{minipage}
\caption{Paired differences in normalized AURC over BO steps 0--18,
using the same methods and paired comparison procedure as in
Figure~\ref{fig:bo-aurc-h10}. Negative values favor the first method.}
\label{fig:bo-aurc-h18}
\end{figure}

Figures~\ref{fig:bo-aurc-h10} and \ref{fig:bo-aurc-h18} give paired AURC
differences for $H=10$ and $18$; negative values favor the first method.
For nested units, all displayed intervals include zero
at $H=10$, and several also include zero at $H=18$. More crossed-unit
comparisons favor F-OMA, particularly against other initializers under
the same GP model. These results do not imply uniform superiority;
intervals are unadjusted for multiple comparisons or horizons.

\begin{figure}[htbp]
\centering
\begin{minipage}[t]{0.49\textwidth}
\centering
\includegraphics[width=\linewidth]
{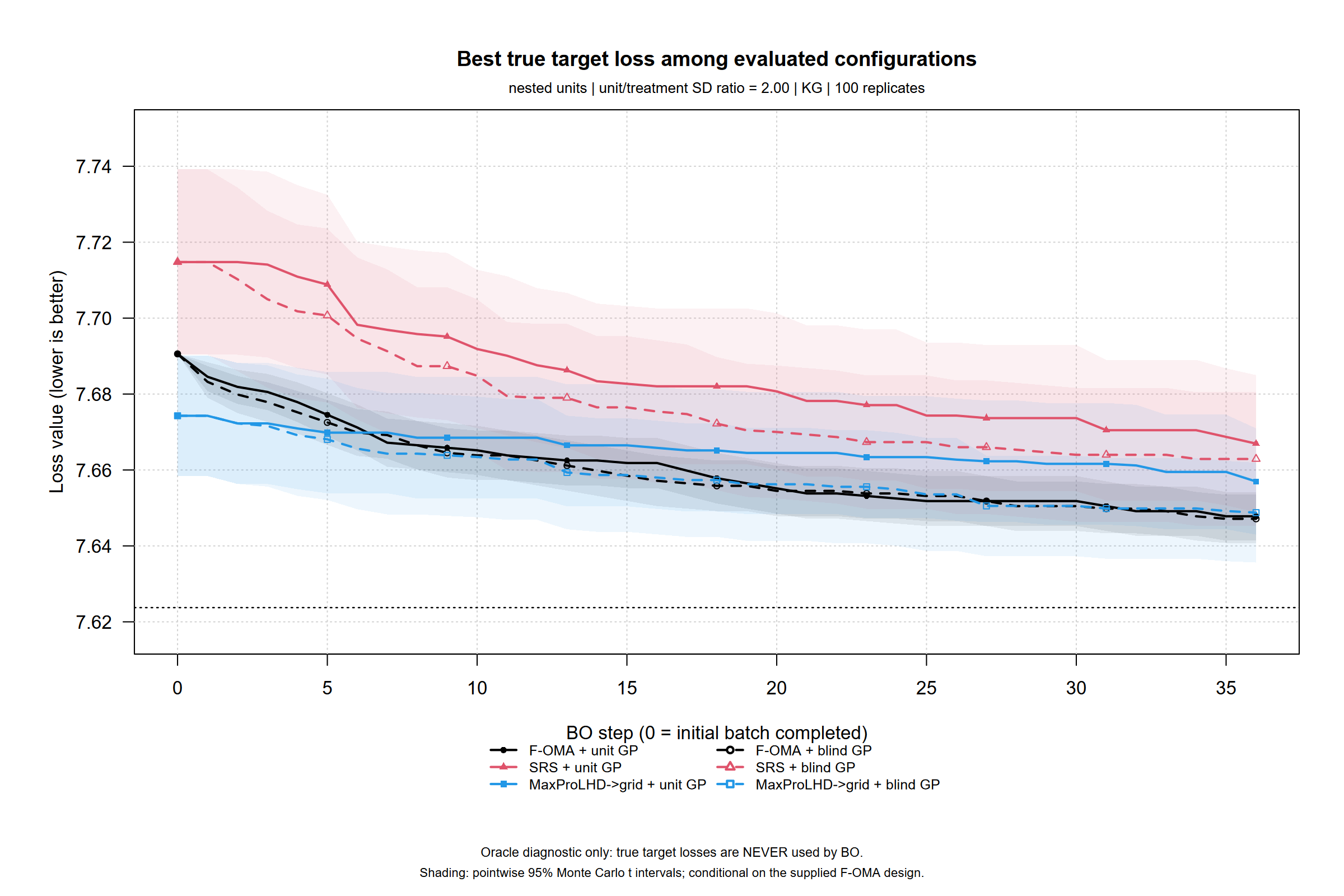}
\smallskip
{\small (a) Nested units}
\end{minipage}
\hfill
\begin{minipage}[t]{0.49\textwidth}
\centering
\includegraphics[width=\linewidth]
{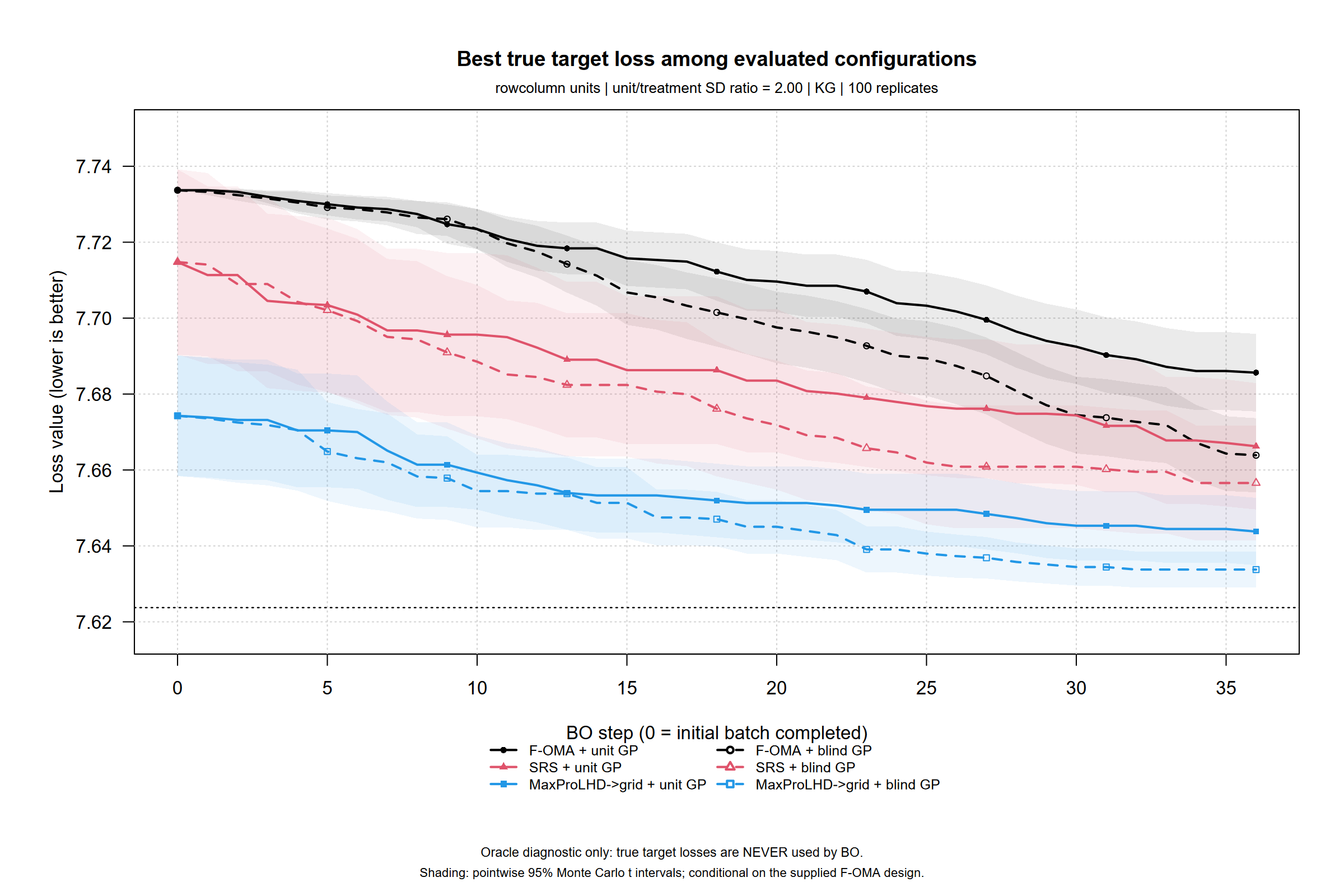}
\smallskip
{\small (b) Crossed units}
\end{minipage}
\caption{Best true target loss among configurations evaluated up to each
BO step. Curves show means over 100 replicates with pointwise 95\%
Monte Carlo intervals. The horizontal line marks the minimum over all
72 candidates. The true target is used only for evaluation, not acquisition
or recommendation. The panel-subtitle ratio compares the RMS of
$g_b(u)+h_b(\mathbf x,u)$ with that of the centered treatment surface,
under the uniform distribution on the complete treatment--unit
product space.}
\label{fig:bo-besttrue}
\end{figure}

\begin{samepage}
\paragraph*{Best-evaluated target loss.}

Figure~\ref{fig:bo-besttrue} reports the lowest true target loss among
configurations evaluated by each step, independently of the GP's
recommendation. The known $R$ is used only for assessment.
For nested units, MaxProLHD$\to$grid starts below F-OMA, with several
methods approaching similar values later; for crossed units, it generally
samples configurations with lower target loss. Sampling a good
configuration therefore need not make it the GP's best recommendation.
\end{samepage}

\end{document}